\documentclass[acmsmall,authordraft,review=false,timestamp=false]{acmart}

\usepackage{graphicx} 
\usepackage{pifont}
\usepackage{xcolor}
\usepackage{amsmath}
\usepackage{graphicx}
\usepackage{caption}
\usepackage{subcaption}
\usepackage{CJKutf8}
\usepackage{enumitem}
\usepackage{multirow}
\usepackage{hyperref}
 
\usepackage{color}
\usepackage[utf8]{inputenc}
\usepackage[switch]{lineno}

\newcommand{\cmark}{\ding{51}}      
\newcommand{\xmark}{\ding{55}}      
\newcommand{\pmark}{$\blacktriangle$}
\AtBeginDocument{%
  }

\setcopyright{acmlicensed}
\copyrightyear{2026}
\acmYear{2026}
\acmDOI{XXXXXXX.XXXXXXX}

\acmJournal{TKDD}

\begin{document}

\title{Eval4DiRec: A Unified and Systematic Evaluation Framework for Diffusion-based Recommender Systems}

\author{Cong Wang}
\orcid{0009-0009-2993-0248}
\affiliation{%
  \institution{Key Laboratory of Computing Power Network and Information Security, Ministry of Education, Shandong Computer Science Center (National Supercomputer Center in Jinan), Qilu University of Technology (Shandong Academy of Sciences); Shandong Provincial Key Laboratory of Computing Power Internet and Service Computing, Shandong Fundamental Research Center for Computer Science}
  \city{Jinan}
  \state{Shandong}
  \country{China}
}
\email{cong.wang2024@gmail.com}

\author{Shoujin Wang}
\orcid{0000-0003-1133-9379}
\affiliation{%
  \institution{University of Technology Sydney}
  \city{Sydney}
  \state{NSW}
  \country{Australia}}
\email{shoujin.wang@uts.edu.au}

\author{Yishuo Li}
\orcid{0009-0002-0093-9854}
\affiliation{%
  \institution{Department of Computer Science and Technology, Tongji University}
  \city{Shanghai}
  \country{China}
}
\email{yishuo.li@foxmail.com}

\author{Qi Zhang}
\affiliation{
  \institution{Department of Computer Science and Technology, Tongji University}
  \city{Shanghai}
  \country{China}
}

\author{Liang Hu}
\affiliation{
  \institution{Department of Computer Science and Technology, Tongji University}
  \city{Shanghai}
  \country{China}
}

\author{Wenpeng Lu}
\orcid{0000-0002-1840-3540}
\authornote{Corresponding author.}
\affiliation{%
  \institution{Key Laboratory of Computing Power Network and Information Security, Ministry of Education, Shandong Computer Science Center (National Supercomputer Center in Jinan), Qilu University of Technology (Shandong Academy of Sciences); Shandong Provincial Key Laboratory of Computing Power Internet and Service Computing, Shandong Fundamental Research Center for Computer Science; Shandong Academy of Artificial Intelligence}
  \city{Jinan}
  \state{Shandong}
  \country{China}
}
\email{wenpeng.lu@qlu.edu.cn}

\renewcommand{\shortauthors}{Wang et al.}

\begin{abstract}
Leveraging the strong generative capabilities and stable training dynamics of diffusion models, diffusion-based recommender systems (RSs) have recently emerged as a novel recommendation paradigm, attracting increasing attention from both academia and industry. However, despite the rapid growth of diffusion-based RSs, a critical issue has emerged: the lack of a unified and systematic quantitative evaluation benchmark, which often results in irreproducible experimental results and unfair comparisons across studies due to inconsistent data processing, training configurations, inference procedures, and evaluation protocols. To address this challenge, we propose Eval4DiRec, the first unified and open-source evaluation framework specifically designed for diffusion-based RSs. Eval4DiRec supports 14 representative diffusion-based RS models across five different recommendation scenarios, providing consistent and reproducible experimental settings to systematically assess their performance. Built upon this framework, we conduct extensive empirical studies to benchmark these models under unified protocols. The results highlight the strong potential of diffusion models for recommendation while also revealing key factors and practical challenges that substantially affect their performance, thereby establishing a solid foundation to facilitate fair evaluation and guide future research in this promising field. Our code and data are available at: https://github.com/wangcong2001/Eval4DiRec.
\end{abstract}




\maketitle

\section{Introduction}

The rapid development of the Internet and the continuous growth of the online user population have resulted in an exponential increase in digital information, bringing significant challenges to data management and information filtering~\cite{dai2021adversarial, wang2024hierarchical, TKDD_survey1, TKDD_survey2, huillingu2026, Wang2020WISE, Wang2022, SpecTD-MR}. To alleviate information overload and improve the user experience, recommender systems (RSs) have become a fundamental component of modern information services. They have been widely adopted across diverse domains such as e-commerce~\cite{ji2021you, qianzhang2025} and streaming media~\cite{goyani2020review}, demonstrating their capability to enhance user engagement and deliver substantial commercial value~\cite{anand2025survey, MR-DTR, RES-MR}. In recent years, the emergence of generative recommender systems (Gen-RSs) has introduced a new paradigm in recommendation research. Unlike traditional discriminative models that directly predict user–item interactions, Gen-RSs aim to learn the underlying data distribution and generate personalized recommendation results, enabling a more accurate and adaptive recommendation process.

Early studies in this direction primarily employed variational autoencoders (VAEs)~\cite{wang2025dr, xiong2025camouflaged} and generative adversarial networks (GANs)~\cite{du2024enhancing} to model user preference generation. These models have shown the capability to dynamically capture the evolution of user preferences~\cite{deldjoo2024review, ramesh2022hierarchical, zhu2023diffusion}. However, despite their promising potential, both paradigms exhibit inherent drawbacks: VAEs often suffer from the issue of posterior collapse~\cite{song2025enhancing}, while GANs are prone to training instability and mode collapse~\cite{becker2022instability, wei2025diffusion}. Such challenges limit their scalability, robustness, and reproducibility in large-scale recommendation scenarios.

\begin{figure}[t]
    \centering
    \begin{subfigure}[b]{0.4\textwidth}
        \includegraphics[width=\linewidth]{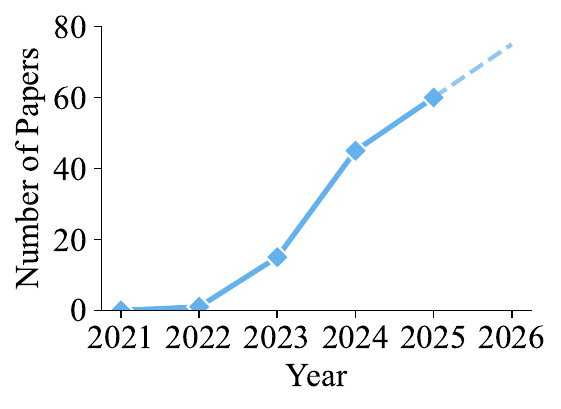}
        \captionsetup{labelfont=normalfont, textfont=normalfont}
        \caption{Paper Count}
        \label{fig:papercount}
    \end{subfigure}
    \begin{subfigure}[b]{0.4\textwidth}
        \includegraphics[width=\linewidth]{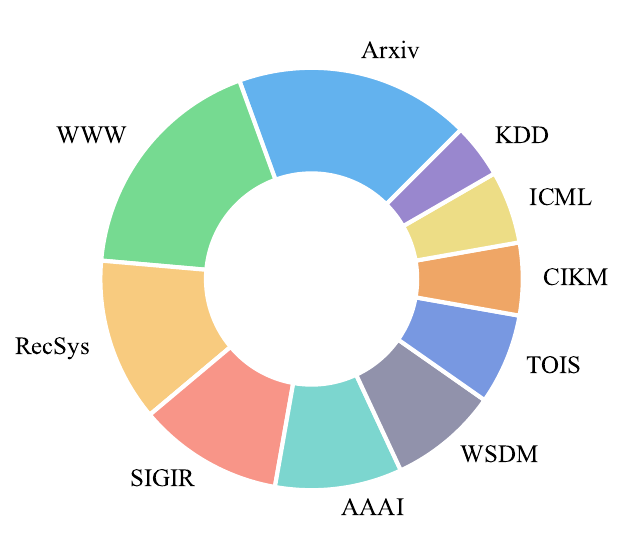}
        \captionsetup{labelfont=normalfont, textfont=normalfont}
        \caption{Paper Distribution}
        \label{fig:paperdistribution}
    \end{subfigure}
    \caption{ (a) The number of papers published annually on diffusion-based recommender systems. (b) The distribution of diffusion-based recommender systems across different venues, showing only the top 10 venues.}
    \label{fig:fig1}
\end{figure}

To address these challenges, two emerging generative paradigms have recently been explored in recommender systems: large language model (LLM)-based RSs and diffusion-based RSs. LLM-based generative RSs leverage the reasoning and knowledge integration capabilities of LLMs to enhance recommendation quality through natural language understanding and generation~\cite{FineMed, liu2026jointcoder, wang2026soaptriage, wang2026rms, zhou2025large,Concept_Representation,WSD_Fanqing,Rumor_zhangqi}. Meanwhile, diffusion-based RSs have shown even greater promise due to their strong probabilistic modeling capacity and training stability~\cite{cao2024survey}. By progressively perturbing and denoising data distributions, diffusion models are capable of capturing complex user–item interaction patterns and generating high-fidelity recommendation signals. 
They have achieved remarkable success across multiple generative domains such as image, text, and video synthesis~\cite{li2023diffusion}, and are now emerging as a transformative foundation for next-generation recommendation architectures~\cite{lee2024stochastic, xia2025s}. 
As illustrated in Figure~\ref{fig:fig1}, diffusion-based RSs have attracted rapidly increasing research attention in recent years.

The application of diffusion models in recommender systems has advanced rapidly, extending across various recommendation tasks and demonstrating their versatility in modeling complex user–item relationships. In particular, diffusion-based RSs have been successfully applied—but are not limited—to the following representative tasks.
\textbf{Collaborative filtering (CF)} remains a fundamental paradigm in recommendation, where user preferences are inferred from the behavior of similar users. In this context, diffusion models are applied to reconstruct and denoise user–item interaction matrices, thereby mitigating sparsity and enhancing representation robustness~\cite{Hyperbolic, ChenCDMR, li2025diffgraph, BuiwwwCold, Zhu_Wang_Wu_2025}. For instance, CODIGEM~\cite{walker2022recommendation} employs a diffusion process to reconstruct user–item interaction vectors, where the test user’s history is diffused forward and subsequently denoised through reverse sampling to generate ranking scores for recommendation.
\textbf{Sequential recommendation} aims to predict the next item a user will interact with based on their historical sequence~\cite{WangwwwUnleashing, luo2025enhancing, ma2024plug, cai2026steering}. 
Diffusion models capture the temporal dynamics of user behaviors by perturbing sequence embeddings and progressively recovering them in reverse, thus modeling long-term dependencies and behavioral uncertainty. A representative example, DiffuRec~\cite{li2023diffurec} is the first work to adapt diffusion models to sequential recommendation by representing items as adaptive distributions rather than fixed vectors, enabling uncertainty-aware modeling of user interests and item aspects.
In \textbf{multimodal recommendation}, diffusion models integrate heterogeneous modalities—such as visual, textual, and acoustic signals—into unified representations to enhance semantic alignment and personalization~\cite{yang2024curricular, ma2025generating, cui2025multi, li2025generating}. For example, DiffMM~\cite{jiang2024diffmm} proposes a multi-modal graph diffusion model that integrates modality-aware graph diffusion with cross-modal contrastive learning to better align multi-modal features with collaborative relations for sparse multi-modal recommendation.
For \textbf{Point-of-interest (POI) recommendation}~\cite{zhang2025survey, long2024diffusion}, diffusion models are used to model spatiotemporal dependencies by learning diffusion trajectories over geographical and temporal embeddings. Diff-POI~\cite{Qin2023diffpoi} introduces a diffusion-based model that captures users’ spatial visiting trends through tailored graph encoders and a diffusion-based sampling strategy to enhance next POI recommendation, particularly in novel geographic areas.
In \textbf{Cross-domain recommendation}~\cite{DiffMSR}, diffusion models facilitate knowledge transfer between rich and cold-start domains through conditional generation and embedding alignment. For instance, DiffCDR~\cite{xuan2024diffusion} diffuses user embeddings in the target domain and conditions on source-domain embeddings in the reverse process, effectively generating informative representations that mitigate negative transfer and enhance cold-start performance.
Overall, these task-specific implementations highlight the adaptability and generative capability of diffusion models in addressing various recommendation challenges, demonstrating their potential as a robust foundation for next-generation recommender systems.

Despite significant advancements in diffusion-based RSs across various tasks and scenarios, the lack of a unified evaluation benchmark remains a critical challenge. 
Existing studies primarily focus on qualitative reviews and comparisons but fail to establish a standardized framework for systematic experimental evaluation for diffusion-based RSs~\cite{lin2024survey, wei2025diffusion}. 
As a result, researchers cannot obtain fair and reliable performance comparisons due to the lack of unified datasets and evaluation protocols, which exacerbates inconsistencies across studies~\cite{li2024rechorus2}. 
These inconsistencies stem from the use of different datasets, diverse data processing strategies, splitting methods, and evaluation metrics. Without standardized evaluation, comparisons become unreliable, ultimately hindering progress in diffusion-based RSs research~\cite{sun2022daisyrec}. 
Although several general recommendation frameworks such as RecBole~\cite{sun2022daisyrec} and ReChorus~\cite{li2024rechorus2} provide standardized environments for evaluating conventional models, they exhibit inherent limitations when applied to diffusion-based RSs. 
This limitation stems from the fundamental paradigm shift between discriminative and generative modeling. 
Unlike traditional recommenders that rely on single-step prediction or embedding optimization to predict user–item interaction scores, diffusion-based recommender systems formulate recommendation as an iterative generative inference process. 
They denoise interaction signals over multiple timesteps and involve stochastic sampling, so the inference process must be explicitly specified via diffusion-related settings. 
In particular, the number of diffusion steps and the noise schedule can systematically affect both the accuracy and diversity of the resulting recommendations. 
Importantly, they fundamentally determine how a diffusion model generates recommendation signals and therefore should be explicitly standardized as part of the evaluation protocol. 
However, existing evaluation frameworks typically do not expose or unify such diffusion-specific configurations in a principled manner, making comparisons across studies easily confounded by inconsistent inference settings. 
In such a context, constructing a robust evaluation framework for fair and reproducible benchmarking has become a critical prerequisite for advancing the research in the field of diffusion-based RSs.

In order to address this significant gap, we propose a novel and the \textit{first} unified \underline{Eval}uation Framework \underline{for} \underline{Di}ffusion-based \underline{Rec}ommendation, named \textbf{Eval4DiRec}.
Eval4DiRec aims to enable fair, reproducible, and comprehensive benchmarking of diffusion-based RS algorithms under consistent experimental protocols, by standardizing data processing and splitting, exposing diffusion-related configurations through a unified interface, and providing consistent training and evaluation pipelines.
The current release supports 14 representative diffusion models across five major recommendation scenarios, serving as a standardized platform for performance evaluation and comparative analysis.
Specifically, Eval4DiRec is composed of four main modules: 
(1) \textbf{Data} module, which loads raw datasets, records basic statistics, performs data splitting, and transforms inputs into a unified format for efficient training and evaluation; 
(2) \textbf{DiRec} module, which integrates diffusion-based recommendation algorithms across five scenarios, and is designed with strong extensibility, allowing easy incorporation of new models for future expansion; 
(3) \textbf{Execution} module, which manages the entire pipeline of model training, validation, and testing; 
(4) \textbf{Utils} module, which provides essential utilities such as configuration parsing, hyperparameter management, logging, and performance evaluation.
Eval4DiRec facilitates reliable benchmarking and thorough analysis of diffusion-based recommendation models under consistent and reproducible experimental settings.

In summary, the work's key contributions are as follows:
\begin{itemize}
    \item We propose Eval4DiRec, the first unified and open-source evaluation framework dedicated to diffusion-based recommender systems, which standardizes the evaluation pipeline, training protocol, inference procedure, and metric computation to enable fair, transparent, and reproducible comparisons.
    \item We implement 14 representative diffusion-based recommender models across five recommendation scenarios within a unified codebase, providing consistent interfaces, configurable components, and an extensible design that facilitates reliable evaluation and future model integration.
    \item We conduct extensive experiments under standardized settings to benchmark diffusion-based recommender systems from multiple perspectives, and further provide systematic analyses on key design choices and robustness under challenging data conditions, yielding empirical findings and practical insights for diffusion-based recommendation research.
\end{itemize}

\section{Related Work}
\label{sec:Analysis}

\subsection{Diffusion Model}

Diffusion models have recently been explored in recommender systems across a variety of recommendation scenarios, including but not limited to collaborative filtering~\cite{jiang2024diffkg, wang2021denoising, li2024recdiff, JingZWYT26}, sequential recommendation~\cite{xie2024breaking, yang2023generate, xie2024bridging, cai2025unleashing}, multimodal recommendation~\cite{cui2025multi, tmmMoLTGL25, ipmKhanS25, inffusXiuT26}, point-of-interest recommendation~\cite{PanZWTWG25, long2024diffusion, tbdZuoYZ26}, and cross-domain recommendation~\cite{eswaJinYLLH25, tsmcDoZZL26, zhao2025distributionally}. These studies suggest that diffusion models can not only serve as generative models, but also serve as denoising, augmentation, representation learning, and sampling mechanisms for recommendation. We summarize several representative scenarios as follows.
\begin{itemize}
    \item \textbf{Collaborative filtering:} Collaborative filtering (CF) remains a fundamental paradigm in recommendation, where user preferences are inferred from the behaviors of similar users~\cite{jiang2024diffkg, wang2021denoising, li2024recdiff, JingZWYT26}. In this context, diffusion models are commonly adopted to reconstruct and denoise sparse user--item interaction signals, such as interaction vectors or matrices, thereby mitigating sparsity and improving the robustness of preference representations. For example, DDRM~\cite{zhao2024denoising} leverages diffusion-style multi-step denoising to robustify user and item embeddings against noisy implicit feedback, guided by collaborative signals and initialized from personalized historical-item representations. In contrast, DiffRec~\cite{wang2023diffusion} performs denoising generation over user interaction vectors and avoids fully corrupting interactions into pure noise to preserve personalization, with extensions such as latent-space diffusion and temporal reweighting for large-scale and dynamic recommendation.
    \item \textbf{Sequential recommendation:} Sequential recommendation aims to predict the next item a user will interact with based on historical interaction sequences~\cite{xie2024breaking, yang2023generate, xie2024bridging, cai2025unleashing}. Diffusion-based methods capture temporal dynamics by perturbing sequence representations and progressively recovering them in the reverse process, which helps model long-range dependencies and behavioral uncertainty. For example, CaDiRec~\cite{cui2024context} proposes a context-aware diffusion model to generate semantically consistent alternative items for specific positions in a user sequence, producing more reasonable augmented views for contrastive learning. By aligning generated items with local context and training the diffusion and recommendation components end-to-end with shared item embeddings, CaDiRec improves sequential recommendation performance compared with random augmentation strategies.
    \item \textbf{Multimodal recommendation:} In multimodal recommendation, diffusion models integrate heterogeneous modalities, such as visual, textual, and acoustic signals, into unified representations to enhance semantic alignment and personalization under sparse interactions~\cite{cui2025multi, tmmMoLTGL25, ipmKhanS25, inffusXiuT26}. For instance, CCDRec~\cite{yang2024curricular} proposes a curriculum-conditioned diffusion framework that aligns multimodal features with collaborative signals and leverages the reverse diffusion process to progressively synthesize and select informative negative samples for multimodal recommendation.
    \item \textbf{Point-of-interest recommendation:} For point-of-interest (POI) recommendation, diffusion models are used to model spatiotemporal dependencies by learning diffusion trajectories over geographical and temporal representations, thereby improving next-POI prediction under complex mobility patterns~\cite{PanZWTWG25, long2024diffusion, tbdZuoYZ26}. Diff-POI~\cite{Qin2023diffpoi} introduces a diffusion-based approach that captures users' spatial visiting trends through tailored graph encoders and a diffusion-based sampling strategy, which is particularly beneficial in novel geographic areas.
    \item \textbf{Cross-domain recommendation:} In cross-domain scenarios, diffusion models facilitate knowledge transfer between data-rich and cold-start domains through conditional generation and embedding alignment, aiming to alleviate data scarcity and reduce negative transfer~\cite{eswaJinYLLH25, tsmcDoZZL26, zhao2025distributionally}. For example, DiffCDR~\cite{xuan2024diffusion} diffuses user embeddings in the target domain and conditions the reverse process on source-domain embeddings, effectively generating informative representations to enhance cold-start performance.
\end{itemize}
Overall, diffusion-based recommender systems differ substantially in their target scenarios, model inputs, conditioning signals, sampling procedures, and evaluation settings. This diversity highlights their potential, but also makes it difficult to compare results across studies under inconsistent experimental protocols. Therefore, a unified evaluation framework is needed to assess diffusion-based recommendation methods fairly and reproducibly across representative scenarios.

\subsection{Survey of Diffusion-based Recommender Systems}

The growing interest in diffusion-based RSs has led to several survey efforts. 
\citet{lin2024survey} were the first to provide a comprehensive review of diffusion models for recommendation, organizing the literature from a pipeline perspective that mirrors real-world recommender system deployment. Their discussion separates prior work into three stages: data engineering, recommender modeling, and content presentation. Under this view, diffusion models are examined according to the functional role they play in each stage, offering an integrated understanding of the emerging field.

\citet{wei2025diffusion} presented a complementary survey that categorizes diffusion-based recommendation methods from the perspective of recommendation scenarios. Their taxonomy emphasizes the recommendation task itself and structures the literature according to the scenario being addressed. This task-oriented viewpoint helps readers understand how diffusion modeling is adapted across diverse recommendation settings.

Beyond survey efforts, \citet{benigni2026diffusion} examined the reproducibility and conceptual suitability of nine DDPM-based recommendation algorithms published at SIGIR 2023 and 2024. Their controlled study reports that only a minority of the original results could be fully reproduced and that well-tuned simpler baselines often outperformed the reported diffusion-based results. It also questions the fit between conventional DDPM generation and Top-$N$ recommendation. Eval4DiRec complements this critical analysis by providing reusable infrastructure for controlled and reproducible comparisons across multiple recommendation scenarios, together with explicit management of diffusion-specific configurations.

Existing surveys~\cite{wei2025diffusion,lin2024survey} organize the field, and the reproducibility analysis of \citet{benigni2026diffusion} establishes the importance of controlled baselines and transparent protocols. A reusable, multi-scenario framework that standardizes data processing, evaluation, and diffusion-specific settings across integrated models nevertheless remains absent. Eval4DiRec addresses this infrastructure gap and evaluates diffusion-based recommendation without assuming its superiority over non-diffusion methods.

\subsection{Evaluation for Recommender Systems}

With the rapid development of RSs, increasing attention has been paid to ensuring reproducibility and fairness in experimental evaluation. In response, several frameworks have been proposed~\cite{meng2022survey, iana2023newsreclib, yang2018openrec, yu2024easyrl4rec, ludewig2018evaluation, lu2025advancing}.

For instance, ReChorus adopts a modular and task-flexible design that supports multiple recommendation scenarios and encourages consistent implementation across tasks~\cite{li2024rechorus2}. RecBole provides a unified framework with extensive model collections and benchmark datasets, together with a configuration-driven workflow that facilitates fair comparisons under consistent settings~\cite{zhao2021recbole}. DaisyRec emphasizes rigorous evaluation by analyzing key hyperparameters and offering standardized evaluation pipelines for reproducible research~\cite{sun2022daisyrec}. In addition, EasyRL4Rec offers a user-friendly library for reinforcement learning-based recommendation with unified evaluation interfaces~\cite{yu2024easyrl4rec}.

Although existing recommendation frameworks provide standardized support for conventional model implementation, training, and evaluation, two important gaps remain when they are applied to diffusion-based recommender systems. \textbf{Gap 1: Lack of Unified Evaluation Protocols.} Existing frameworks provide only limited or model-specific support for diffusion-based models. Consequently, these models are often evaluated using different datasets, data processing strategies, splitting methods, training procedures, and metric implementations, making the reported results difficult to compare fairly and reliably. \textbf{Gap 2: Lack of Standardized Diffusion-Specific Inference Configurations.} Unlike conventional recommenders based mainly on single-step prediction, diffusion-based recommenders rely on iterative denoising and stochastic sampling. Their performance can be substantially affected by settings such as the number of diffusion steps, noise schedule, noise distribution, and reverse-sampling procedure.

However, existing frameworks generally embed these settings within individual model implementations rather than exposing and standardizing them as part of a shared evaluation protocol. These gaps cannot be addressed simply by adding individual diffusion-based models to an existing library, because fair benchmarking requires both consistent general evaluation protocols and unified management of diffusion-specific configurations across models and recommendation scenarios. As summarized in Table~\ref{tab:framework_comparison}, Eval4DiRec addresses these gaps by standardizing data processing, splitting, training, inference, and metric computation, while exposing diffusion-process and reverse-sampling configurations as framework-level components for 14 models across five recommendation scenarios.

\begin{table*}[t]
\centering
\begingroup
\color{black}
\caption{
Comparison of representative recommendation frameworks and evaluation studies.
\cmark, \pmark, and \xmark denote native, partial or model-specific, and no explicitly documented support, respectively.
CF, Seq., MM, POI, and CDR denote collaborative filtering, sequential, multimodal, point-of-interest, and cross-domain recommendation.
Diffusion-aware support covers process configuration, sampling protocols, and diagnostic analyses.
}
\label{tab:framework_comparison}
\resizebox{\textwidth}{!}{
\begin{tabular}{lcccccccc}
\toprule
\multirow{2}{*}{\textbf{Framework}}
& \multicolumn{5}{c}{\textbf{Recommendation Scenario Coverage}}
& \multicolumn{3}{c}{\textbf{Diffusion-aware Support}}
\\
\cmidrule(lr){2-6}
\cmidrule(lr){7-9}
& CF
& Seq.
& MM
& POI
& CDR
& Process Config
& Sampling Protocol
& Diffusion Diagnostics
\\
\midrule
ReChorus~\cite{li2024rechorus2}
& \cmark & \cmark & \xmark & \xmark & \xmark
& \xmark & \xmark & \xmark
\\
DaisyRec~\cite{sun2022daisyrec}
& \cmark & \xmark & \xmark & \xmark & \xmark
& \xmark & \xmark & \xmark
\\
RecBole~\cite{zhao2021recbole}
& \cmark & \cmark & \cmark & \xmark & \cmark
& \pmark & \pmark & \xmark
\\
Ludewig \& Jannach~\cite{ludewig2018evaluation}
& \cmark & \cmark & \xmark & \xmark & \xmark
& \xmark & \xmark & \xmark
\\
NewsRecLib~\cite{iana2023newsreclib}
& \xmark & \cmark & \cmark & \xmark & \xmark
& \xmark & \xmark & \xmark
\\
OpenRec~\cite{yang2018openrec}
& \cmark & \cmark & \cmark & \xmark & \xmark
& \xmark & \xmark & \xmark
\\
EasyRL4Rec~\cite{yu2024easyrl4rec}
& \cmark & \cmark & \xmark & \xmark & \xmark
& \xmark & \xmark & \xmark
\\
\midrule
\textbf{Eval4DiRec}
& \cmark & \cmark & \cmark & \cmark & \cmark
& \cmark & \cmark & \cmark
\\
\bottomrule
\end{tabular}
}
\endgroup
\end{table*}

\section{Preliminaries of Diffusion Models}
\label{sec:Preliminary}

Diffusion models have recently emerged as a powerful generative paradigm for recommendation, showing strong capability in modeling complex data distributions via forward diffusion process and reverse denoising process~\cite{li2023diffusionx, diffusion_tkdd1}. For completeness, we summarize diffusion modeling under two commonly used formulations: denoising diffusion probabilistic models (DDPMs) and score-based diffusion models (SDEs). 

\subsection{Denoising Diffusion Probabilistic Models (DDPMs)}

As the most typical type of diffusion models, Denoising Diffusion Probabilistic Models (DDPMs)~\cite{ho2020denoising} are established on a well-defined probabilistic framework, where the forward diffusion and reverse denoising processes operate as two Markov chains. In the forward process, predefined noise is progressively injected into data samples through a Markovian transition kernel, while the reverse process employs neural networks to reconstruct the original data distribution by iteratively eliminating the injected noise.

\textbf{Forward diffusion process}. Forward diffusion process involves sampling initial data $x_0$ from a given data distribution $q(x)$, and gradually adding noise to obtain a pure noise sample $x_T$. The noise addition is controlled by a variance schedule \{$\beta_1,\beta_2,\cdots,\beta_T$\}, which follows predefined schedules~\cite{hang2024improved, wang2023learning} such as linear or cosine to regulate the noise injection magnitude. Formally, this process is modeled as a Markov chain where each transition $x_{t-1}$ to $x_t$, introduces Gaussian noise with variance $\beta_t$, progressively corrupting the data until it becomes indistinguishable from random noise. The transition kernels are defined as:
\begin{equation}
\begin{aligned}
    q(x_t|x_{t-1})= \mathcal{N}(x_t;\sqrt{1-\beta_t}x_{t-1}, \beta_t\mathbf{I}),
\end{aligned}
\end{equation}
where $\mathbf{I}$ denotes the identity matrix with the same dimensionality as $x_{t-1}$, and $\mathcal{N}(x;\mu, \sigma\mathbf{I})$ represents the Gaussian distribution of $x$ with mean $\mu$ and covariance matrix $\sigma\mathbf{I}$. We can directly obtain $x_t$ from $x_0$ through a series of transition kernels, as shown below: 
\begin{equation}
\begin{aligned}
    q(x_t|x_{0}) &= \prod_{k=1}^{t}q(x_k|x_{k-1}) \\
                 &= \mathcal{N}(x_t;\sqrt{\bar{\alpha}_t}x_{0}, \sqrt{1-\bar{\alpha}_t}\mathbf{I}),
\end{aligned}
\end{equation}
where  $\alpha_t=1-\beta_t$ and $\bar{\alpha}_t = \prod_{i=1}^t\alpha_i$. 

\textbf{Reverse denoising process}. Reverse denoising process iteratively reconstructs the original data sample $x_0$ through a Markov chain. Specifically, the process begins by sampling initial noise $x_T$  from a standard Gaussian distribution  $\mathcal{N}(\mathbf{0}, \mathbf{I})$, and then generates data samples through learnable Gaussian kernels $p_\theta(x_{t-1}|x_t)$:
\begin{equation}
\begin{aligned}
    p_\theta(x_{t-1}|x_t) = \mathcal{N}(x_{t-1};\mu_\theta(x_t,t), \sigma_\theta(x_t,t)\mathbf{I}),
\end{aligned}
\end{equation}
where $\mu_\theta$ and $\sigma_\theta$ are learnable mean and variance of the reverse Gaussian kernels.

\textbf{Training}. The training objective of diffusion models is to optimize the negative log-likelihood by minimizing the variational lower bound. The loss function consists of three components: the prior loss $L_T$, the reconstruction loss $L_0$, and the KL divergence loss $L_{t-1}$ at intermediate timesteps as follows:
\begin{equation}
\begin{aligned}
    \mathbb{E}[-\log p_\theta(x_0)] &\le \mathbb{E}_q[\underbrace{D_{KL}(q(x_T|x_0)||p(x_T))}_{L_T}\\
    &+ \sum_{t>1}\underbrace{D_{KL}(q(x_{t-1}| x_t, x_0) || p_\theta(x_{t-1}|x_t))}_{L_{t-1}}  \\
    &- \underbrace{\log p_\theta(x_0|x_1)}_{L_0}].
\end{aligned}
\end{equation}

$L_T$ is a prior loss that is minimized when the final latent distribution matches the Gaussian prior; it requires no optimization as it has no trainable parameters.
$L_0$ is used to predict the log probability of the original data sample given the first-step latent.
$L_{t-1}$ ensures that the intermediate distribution remains consistent between the forward and backward processes~\cite{luo2022understanding}.
For simplification~\cite{ho2020denoising}, it can also be formulated as:
\begin{equation}
\begin{aligned}
    L_{simple} &= \mathbb{E}_{t,x_0,\epsilon}
    \left[ 
    {\lVert
    \epsilon 
    - \epsilon_\theta(\sqrt{\bar{\alpha}_t}x_0 
    + \sqrt{1 - \bar{\alpha}_t}\epsilon,t) 
    \rVert}^2
    \right]. \\
\end{aligned}
\end{equation}

\textbf{Inference}. Given noisy data $x_T$, we iteratively generate the data $x_0$ through $T$ steps reverse denoising process, as follows:
\begin{equation}
\begin{aligned}
    p_\theta (x_{t-1} | x_t) &= \mathcal{N}(x_{t-1};\mu_\theta(x_t, t), \sigma_\theta(x_t, t)\mathbf{I}) \\
                             &= \frac{1}{\sqrt{\alpha_t}}(x_t - \frac{\beta_t}{\sqrt{1-\bar{\alpha}_t}}\epsilon_\theta(x_t, t)) + \sigma_\theta(x_t, t)\textbf{z},
\end{aligned}
\end{equation}
where $z \sim \mathcal{N}(\mathbf{0}, \mathbf{I})$, and $\beta_t \approx \sigma^2_\theta$.

\subsection{Score-based Diffusion Models (SDEs)}
Score-based diffusion models further generalize the discrete-time diffusion process in DDPMs to a continuous-time framework based on stochastic differential equations (SDEs)~\cite{SDE}. Instead of a finite sequence $\{x_0,\ldots,x_T\}$, the diffusion process is modeled as a continuous-time stochastic process $x(t)$ with $t\in[0,T]$. 

\textbf{Forward diffusion process}. 
The forward diffusion process is defined by a SDE: 
\begin{equation}
\begin{aligned}
    \mathrm{d}x = f(x,t)\mathrm{d}t + g(t)\mathrm{d}w, \quad t\in[0,T],
\end{aligned}
\end{equation}
where $f(x,t)$ is the drift term, $g(t)$ is the diffusion coefficient, and $w$ denotes the standard Wiener process. 
A commonly used instantiation is the variance-preserving SDE (VP-SDE), which can be viewed as the continuous-time counterpart of DDPMs:
\begin{equation}
\begin{aligned}
    \mathrm{d}x = -\frac{1}{2}\beta(t)x\,\mathrm{d}t + \sqrt{\beta(t)}\,\mathrm{d}w.
\end{aligned}
\end{equation}

With an appropriate schedule $\beta(t)$, the forward process gradually transforms the data distribution into a simple prior distribution.

\textbf{Reverse denoising process}. 
Sampling is performed by solving the reverse-time SDE associated with the forward process:
\begin{equation}
\begin{aligned}
    \mathrm{d}x = \left[f(x,t) - g^2(t)\nabla_x \log p_t(x)\right]\mathrm{d}t + g(t)\mathrm{d}\bar{w},
\end{aligned}
\end{equation}
where $\bar{w}$ is the reverse-time Wiener process and $p_t(x)$ is the marginal density of $x(t)$. 
The term $s^\ast(x,t)=\nabla_x \log p_t(x)$ is referred to as the \emph{score function}. 
In practice, a neural network $s_\theta(x,t)$ is trained to approximate $s^\ast(x,t)$, enabling the reverse-time dynamics. 
Given the forward SDE, the score network can be learned via denoising score matching by minimizing:
\begin{equation}
\begin{aligned}
    \mathcal{L}
    = \mathbb{E}_{t, x_0, x_t}
    \left[\lambda(t)\left\lVert s_\theta(x_t,t) - \nabla_{x_t}\log p(x_t|x_0)\right\rVert_2^2\right],
\end{aligned}
\end{equation}
where $t\in[0,T]$ is sampled from a predefined distribution, $x_0$ is sampled from the data distribution, and $x_t$ is obtained by running the forward diffusion process up to time $t$, i.e., $x_t \sim p(\cdot \mid x_0)$. 
$\lambda(t)$ is a positive weighting function to balance the contributions of different time regions during optimization. 
For many commonly used SDEs, the forward transition density $p(x_t\mid x_0)$ is Gaussian, making $\nabla_{x_t}\log p(x_t\mid x_0)$ tractable. 
By learning $s_\theta(x_t,t)$ in this way, we avoid directly estimating the intractable score $\nabla_{x_t}\log p_t(x_t)$ of the unknown marginal distribution.

\section{Eval4DiRec}
\label{sec:Eval4DiRec}

\begin{figure}[tb]
    \centering
    \includegraphics[width=0.95\linewidth]{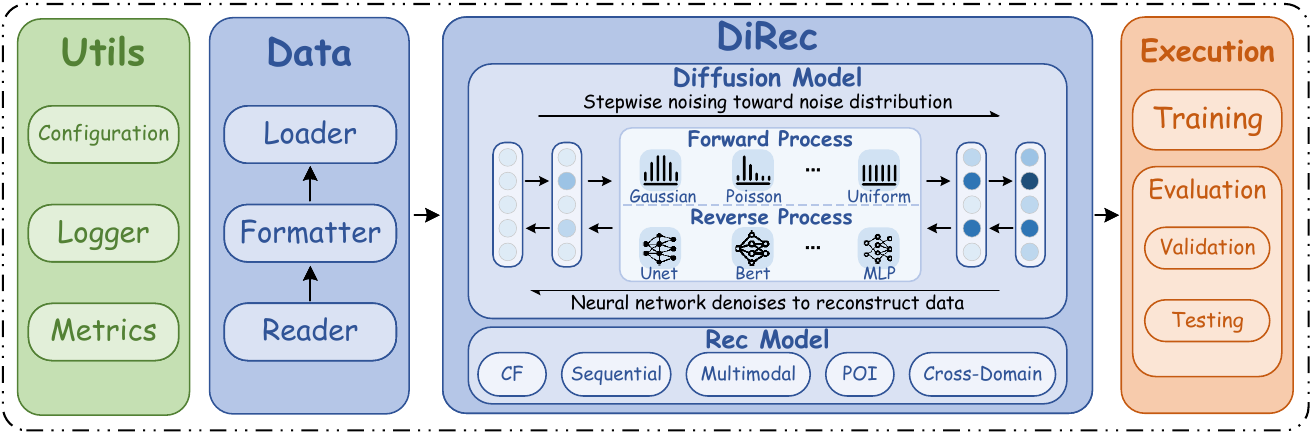}
    \caption{Overview of Eval4DiRec. The framework standardizes the end-to-end evaluation pipeline for diffusion-based recommender systems, including data preprocessing/splitting, model integration, training and inference, and metric reporting. It consists of four modules: \textit{Data} prepares unified inputs and scenario-specific splits; \textit{DiRec} integrates diffusion-based recommenders and manages sampling-related configurations; \textit{Execution} orchestrates training, validation, and testing with structured outputs; and \textit{Utils} provides utilities for configuration, randomness control, logging, and metric computation to support fair and reproducible comparisons.}
    \label{fig:Framework}
\end{figure}

Eval4DiRec is a unified evaluation framework tailored for diffusion-based recommender systems. As illustrated in Figure~\ref{fig:Framework}, it standardizes the end-to-end experimental pipeline, including data processing, model training and inference, and evaluation protocols, to minimize comparison bias caused by inconsistent dataset splits, preprocessing choices, metric implementations, and sampling configurations. The current release supports 14 representative methods spanning five recommendation scenarios, providing a reproducible benchmark for future research. 
Eval4DiRec is composed of four modules. 
The \textit{Data} module converts raw datasets into a unified, trainable format, including basic statistics reporting, basic preprocessing and standardization, and scenario-specific splitting and sample construction. 
The \textit{DiRec} module integrates diffusion-based recommendation algorithms through a standardized interface and an extensible registry, and treats the sampling process as an explicit, configurable component so that the number of steps, noise schedule, and sampling-related settings are consistently managed and recorded.
The \textit{Execution} module orchestrates a standardized workflow for training, validation, and testing, supports consistent early stopping and tuning routines, and outputs structured logs and evaluation results. 
The \textit{Utils} module provides essential utilities for configuration management, randomness control, logging, and metric computation, ensuring that all methods are compared under the same evaluation implementation.

Figure~\ref{fig:Framework} presents the workflow and input--output contracts of Eval4DiRec. 
The Data module processes raw interaction records and optional side information into scenario-aware train/validation/test loaders and records metadata such as the numbers of users and items and the feature dimensions. 
The DiRec module consumes these standardized batches and instantiates the recommendation backbone and diffusion process from a shared configuration. 
The Execution module constructs the data and model objects, runs training, selects checkpoints using validation metrics, evaluates the selected model on the test split, and saves logs and metadata. 
The Utils module supplies configuration parsing, registry lookup, random seed initialization, metric computation, and result logging across the other modules. 
Together, these contracts map raw datasets to standardized batches, training losses, ranking scores, and reproducible evaluation outputs.

To ensure fairness and reproducibility, Eval4DiRec standardizes key factors that can affect evaluation outcomes, including data preprocessing and splitting rules, training and validation procedures, metric implementations, random seed control, and inference configurations. The framework outputs standardized experiment configurations and logs, enabling results to be fully traceable and reproducible while reducing uncertainty caused by environment differences and implementation details. Meanwhile, its modular design lowers the cost of extension, allowing new models or datasets to be integrated by implementing unified interfaces and providing configuration files without altering the established evaluation protocol.

\subsection{Data Module}

The data module is designed to provide flexible, scalable, and consistent data processing across diverse recommendation tasks. 
Its primary goal is to reduce the integration cost of new datasets by adopting a simple yet highly adaptable input format, thereby facilitating both reproducibility and extensibility.
Specifically, it transforms raw datasets into a unified structure optimized for efficient training and evaluation. 
During this transformation, it automatically computes and logs dataset statistics such as interaction sparsity and the number of users and items, which are instrumental for dataset analysis, model diagnostics, and reproducibility. 
To accommodate different recommendation settings, the module offers customized processing pipelines tailored to five recommendation scenarios. 
Each scenario is equipped with a standardized input specification to ensure consistency and compatibility across models and tasks. 
Following preprocessing, the module produces standardized data loaders dedicated to training, validation, and testing phases, ensuring seamless integration into the model execution pipeline.
With minimal configuration, users can flexibly incorporate new datasets into existing workflows, making the framework suitable not only for replicating prior research but also for enabling future extensions.

More concretely, the customized processing pipelines refer to the scenario-specific procedures that convert raw files into the supervision signals required by each task. 
For collaborative filtering, the pipeline reads user-item interaction records, remaps raw identifiers to contiguous indices, constructs train/validation/test interactions, and provides candidate item sets for Top-$K$ ranking. 
For sequential recommendation, the pipeline additionally sorts interactions by timestamp and constructs prefix-target samples under the leave-one-out protocol. 
For multimodal recommendation, the pipeline aligns interaction records with item-side visual, textual, or acoustic features and checks that feature tensors are indexed consistently with item identifiers. 
For POI recommendation, the pipeline preserves temporal and spatial check-in information so that models can use mobility-related signals. 
For cross-domain recommendation, the pipeline loads source-domain and target-domain interactions, keeps the user and item spaces for each domain, and prepares the transfer setting used by the corresponding model. 
Although preprocessing and sample construction follow the assumptions of each scenario, all pipelines expose a common training, validation, and testing interface to the downstream Execution module.

The standardized input specification defines the data contract between the Data module and the remaining components of Eval4DiRec.
At the file level, processed interaction records are organized using user and item identifiers, together with scenario-dependent attributes such as timestamps, ratings, domain labels, POI coordinates, and paths to modality features.
At runtime, scenario-specific data loaders transform these records into batches with consistent field names and tensor structures, including user and item identifiers, historical interaction sequences, target items, candidate or negative items, masks, and optional multimodal or spatiotemporal features.
This design enables models within the same recommendation scenario to share the same processed data and evaluation protocol, while methods requiring specialized sample construction can be supported through registered data-processing profiles.
Concrete input examples and configuration details are available in the public repository.
The grouped YAML files under \texttt{configs/} specify dataset paths, recommendation scenarios, data splitters, and data loaders, while \texttt{eval4direc/data/} contains the corresponding data contracts and scenario-specific loader implementations.

\subsubsection{Datasets}
\label{appendix:datasets}

Table~\ref{tab:datasets} reports the statistics of the processed datasets used in Eval4DiRec, rather than the raw dataset statistics. 
Before model training and evaluation, all datasets are converted into unified implicit-feedback recommendation data through scenario-specific preprocessing, including rating-threshold filtering, duplicate interaction handling, user/item filtering, and train--validation--test splitting. 
Therefore, the reported number of interactions may differ from the original dataset size. 
For example, ML-1M originally contains about one million explicit ratings, while the processed version used in our experiments contains 223,456 implicit interactions after applying the unified preprocessing protocol. 
Reporting processed statistics ensures that the dataset scale is consistent with the actual data used by all evaluated models.

\textbf{Collaborative Filtering and Sequential}:
\textbf{Yelp}
\footnote{\url{https://www.yelp.com/dataset}}
is released by Yelp as a public collection of real-world data from its platform. 
It includes information about businesses and users, along with user-generated content such as reviews, ratings, photos, and check-ins.
\textbf{Beauty} is a subset of the \textit{Amazon review data}
\footnote{\url{https://cseweb.ucsd.edu/~jmcauley/datasets/amazon/links.html}}
~\cite{he2016ups, mcauley2015image}, collected from the Amazon e-commerce platform. 
The \textit{Amazon review data} is a large-scale public corpus of user-generated product reviews and ratings on Amazon, accompanied by rich item metadata such as titles, categories, and related-product links.
\textbf{MovieLens}~\cite{harper2015movielens} is released by the GroupLens research group and contains user-provided movie ratings and related metadata collected from the MovieLens website.
\textbf{ML-1M} \footnote{\url{https://grouplens.org/datasets/movielens/1m/}} is the most widely used version which contains 1M movie ratings.

\textbf{Multimodal}:
\textbf{TikTok}~\cite{jiang2024diffmm} is a multimodal dataset built from user interactions with short-video content, where each item is associated with visual, acoustic, and textual information. 
The textual modality is typically derived from video captions and related user-generated text, providing complementary semantic signals alongside the video and audio features.
\textbf{Baby} and \textbf{Sports} are category-specific subsets of the \textit{Amazon review data} that pair user-item interactions with multimodal item content. Each product is associated with textual descriptions and product images, enabling the use of both textual and visual modalities in addition to interaction data.

\begin{table}[t]
    \centering 
    \caption{Statistics of processed datasets used in Eval4DiRec}
    \label{tab:datasets}
    \begin{tabular}{l| c c c c}
        \hline
        \textbf{Dataset}         & \textbf{\#User}    & \textbf{\#Item}    & \textbf{\#Interactions}  & \textbf{Sparsity}  \\
        \hline
        \textbf{Yelp}       & 37,232    & 47,744    & 359,330   & 99.98\% \\
        \textbf{Beauty}     & 9,895     & 8,455     & 73,528    & 99.91\%   \\
        \textbf{ML-1M}           & 5,652    & 2,468    & 223,456      & 98.40\%   \\
        \hline
        \textbf{TikTok}          & 9,319    & 6,710    & 59,541       & 99.90\%   \\
        \textbf{Baby}     & 19,445    & 7,050      & 139,110     & 99.89\%   \\
        \textbf{Sports}   & 35,598       & 18,357     & 256,308        & 99.96\%   \\
        \hline
        \textbf{NYC}             & 1,083   & 9,989   & 179,468      & 98.34\%   \\
        \textbf{TKY}             &  2,293   & 15,177    & 494,807      & 98.58\%   \\
        \textbf{Gowalla}         & 10,162     & 24,237    & 456,820      & 99.81\%   \\
        \hline
        \textbf{Video}           & 123,960   & 50,052    & 1,697,533      & 99.97\%   \\
        \textbf{Music}           & 75,258     & 64,443     & 1,097,592        & 99.98\%   \\
        \textbf{Book}            & 603,668    & 367,982    & 8,898,041       & 99.99\%   \\
        \hline
    \end{tabular}
\end{table}

\textbf{Point of Interest}:
\textbf{TKY} (Tokyo) and \textbf{NYC} (New York City) are subsets of a public \textit{Foursquare}\footnote{\url{https://sites.google.com/site/yangdingqi/home/foursquare-dataset}} check-in collection, which records users' visits to venues on the platform~\cite{yang2015nationtelescope, yang2016participatory}. 
Each record typically consists of user and venue identifiers along with temporal information, providing a real-world trace of user mobility and venue visits.
\textbf{Gowalla}\footnote{\url{https://snap.stanford.edu/data/loc-gowalla.html}} is a public check-in dataset released by SNAP, containing location-sharing traces from the Gowalla platform. 
It provides user check-in records to locations, and is often accompanied by the corresponding friendship network information from the same platform.

\textbf{Cross-domain}:
Following previous works~\cite{zhu2022personalized}, we use three real-world category subsets from the \textit{Amazon review data}, namely \textbf{Video} (Movies and TV), \textbf{Music} (CDs and Vinyl), and \textbf{Book} (Books), each of which contains user-generated reviews and ratings together with associated item metadata. 
Based on these three domains, we construct cross-domain recommendation tasks by treating one domain as the source and another as the target, and evaluate domain transfer under three settings: Video$\rightarrow$Music, Book$\rightarrow$Video, and Book$\rightarrow$Music~\cite{zhu2022personalized}.

To ensure consistency in evaluation, all datasets are first processed under scenario-specific but unified preprocessing protocols and are subsequently split according to the corresponding recommendation setting.
Collaborative filtering and sequential recommendation share the same preprocessing pipeline. For the explicit-feedback datasets used in these two scenarios, we first convert ratings into positive implicit-feedback interactions by retaining only interactions with a rating of 5. We then apply 5-core filtering, ensuring that every retained user and item is associated with at least five interactions. After filtering, the interactions of each user are arranged in chronological order, and the original user and item identifiers are remapped to consecutive integer indices.
For collaborative filtering, each user's chronologically ordered interaction history is partitioned into training, validation, and test sets using a 7:1:2 ratio.
For sequential recommendation, we adopt a leave-one-out protocol: the last interaction in each user's chronological sequence is held out for testing, the second-to-last interaction is used for validation, and all preceding interactions are used for training. This design ensures that model training and validation rely only on interactions that occur before the test interaction, thereby preventing future information from leaking into the training process.
For multimodal recommendation~\cite{jiang2024diffmm, ma2024multimodal}, we align item-side features with the processed interaction data using consistent item indices. TikTok provides 128-dimensional visual, 128-dimensional acoustic, and 768-dimensional textual features, whereas Amazon Baby and Sports provide 4,096-dimensional visual and 1,024-dimensional textual features. All modalities are stored in their original feature spaces, with dimensional alignment and multimodal fusion performed by model-specific projection layers. All multimodal models share the same processed user--item interactions, item-index mapping, and candidate sets to ensure fair comparison.
For POI recommendation~\cite{Qin2023diffpoi}, we apply 5-core filtering to remove users and POIs with fewer than five check-ins while retaining timestamps and geographical coordinates. Each user's check-ins are chronologically divided into training, validation, and test sets using a 7:1:2 ratio on the Tokyo, New York City, and Gowalla datasets, thereby preventing future-information leakage. The validation set is used for model selection, and the selected checkpoint is evaluated on the test set. All POIs are considered as candidates during evaluation instead of sampled negatives.
For cross-domain recommendation~\cite{xuan2024diffusion}, we preserve separate source- and target-domain item spaces and construct the Video$\rightarrow$Music, Book$\rightarrow$Video, and Book$\rightarrow$Music tasks using overlapping users. These users are divided into training, validation, and test sets with a 7:1:2 ratio. We further convert explicit ratings into implicit interactions and evaluate target-domain predictions as a Top-$K$ ranking task.
The code and datasets are available at https://github.com/wangcong2001/Eval4DiRec.

\subsection{DiRec Module}

The DiRec module is the core component of Eval4DiRec, responsible for integrating and evaluating diffusion-based recommendation algorithms under a unified protocol. 
It consists of two submodules: \textit{Rec Model} and \textit{Diffusion Model}. 
The \textit{Rec Model} defines the backbone recommendation architecture and representation learning strategy, such as user-item embedding learning, sequential encoding, or multimodal fusion, while the \textit{Diffusion Model} implements the key mechanisms of the diffusion process in the recommendation architecture, including the forward process, reverse process, and the noise schedule. 
These components jointly determine the model's sampling process and can substantially affect the quality and stability of the resulting recommendation signals.
By decoupling the recommendation backbone from the diffusion process, DiRec enables a unified training, sampling, and evaluation workflow across different recommendation scenarios, and unifies diffusion configurations and sampling procedures via standardized interfaces, facilitating consistent integration of new models and continuous expansion of the model library. 

In practice, this module is driven by the \textit{Rec Model} and \textit{Diffusion Model} sections of a grouped YAML configuration. 
The \textit{Rec Model} section selects a registered recommender and its hyperparameters, while the \textit{Diffusion Model} section specifies the number of training timesteps, inference steps, noise schedule, noise distribution, prediction type, sampler, and process-specific parameters. 
The registry then resolves these declarations into a recommender object and a diffusion process object. 
The standardized model interface requires each recommender to expose a training-loss function and an inference function that produces ranking scores, so the execution module does not need method-specific training scripts. 
Meanwhile, diffusion schedules, samplers, noise distributions, and process implementations are separately registered, which makes it possible to compare sampling choices or add a new diffusion variant without rewriting the recommender backbone. 
In our current release, we have systematically collected and reviewed nearly 90 relevant research papers and implemented 14 representative diffusion-based recommendation models across 5 distinct recommendation scenarios, with T-DiffRec treated as a time-aware sequential variant. 
Our implementation covers both recently proposed models (e.g., CCDRec~\cite{yang2024curricular}) and earlier foundational baselines that have significantly shaped the diffusion-based recommendation paradigm (e.g., CODIGEM~\cite{walker2022recommendation}, DiffRec~\cite{wang2023diffusion}).

\subsubsection{Supported Models}

The implemented models were selected to provide representative rather than exhaustive coverage of diffusion-based recommendation research. Specifically, we considered scenario coverage, methodological diversity, research development, and reproducibility. The benchmark includes 14 models spanning five major recommendation scenarios, namely collaborative filtering, sequential, multimodal, point-of-interest, and cross-domain recommendation. These methods cover diverse uses of diffusion, including interaction reconstruction, representation denoising, conditional generation, sequence augmentation, multimodal graph modeling, and cross-domain transfer. By including both foundational approaches and recent task-specific developments with sufficiently detailed descriptions or publicly available implementations, the benchmark suite captures the main research directions of the field while remaining suitable for unified evaluation. To make the coverage and representativeness of the selected models more explicit, Table~\ref{tab:selected_models} summarizes the 14 models according to their recommendation scenarios and the main reasons for their inclusion.

\begin{table*}[t]
\renewcommand{\arraystretch}{0.9}
\setlength{\tabcolsep}{1pt} 
\centering
\scriptsize
\caption{Summary and selection rationale of the 14 diffusion-based recommendation models implemented in Eval4DiRec.}
\label{tab:selected_models}
\begin{tabular}{lll}
\toprule
\textbf{Scenario}
& \textbf{Model}
& \textbf{Reason for Selection} \\
\midrule

\multirow{4}{*}{\centering
Collaborative Filtering}
& CODIGEM~\cite{walker2022recommendation}
& An early diffusion-based model for reconstructing user--item interaction vectors. \\

& DiffRec~\cite{wang2023diffusion}
& A representative interaction-vector diffusion model for collaborative filtering. \\

& CF-Diff~\cite{hou2024collaborative}
& Represents the use of high-order collaborative signals in diffusion-based RSs. \\

& DDRM~\cite{zhao2024denoising}
& Represents diffusion-based denoising of user and item representations. \\

\midrule

\multirow{5}{*}{\centering
Sequential Recommendation}
& DiffuRec~\cite{li2023diffurec}
& An early diffusion-based model designed for sequential recommendation. \\

& DiffuASR~\cite{liu2023diffusion}
& Represents the use of diffusion models for sequence augmentation. \\

& T-DiffRec~\cite{wang2023diffusion}
& Represents time-aware interaction modeling in diffusion-based RSs. \\

& CaDiRec~\cite{cui2024context}
& Represents context-aware diffusion augmentation for sequential recommendation. \\

& CDDRec~\cite{wang2024conditional}
& Represents conditional diffusion guided by historical interaction sequences. \\

\midrule

\multirow{3}{*}{\centering Multimodal Recommendation}
& MCDRec~\cite{ma2024multimodal}
& Represents multimodal representation learning and graph denoising. \\

& DiffMM~\cite{jiang2024diffmm}
& Represents diffusion-based multimodal user--item graph generation. \\

& CCDRec~\cite{yang2024curricular}
& Represents multimodal alignment and adaptive negative sampling. \\

\midrule

POI Recommendation
& Diff-POI~\cite{Qin2023diffpoi}
& Provides representative coverage of diffusion-based POI recommendation. \\

\midrule

Cross-Domain Recommendation
& DiffCDR~\cite{xuan2024diffusion}
& Provides representative coverage of diffusion-based cross-domain knowledge transfer. \\

\bottomrule
\end{tabular}
\end{table*}

For models with publicly available official implementations, we used their official repositories as the primary references. The integrated versions preserve the core model architectures. 
To enable unified evaluation, we standardized noise schedules, diffusion steps, noise distributions, and sampling configurations, and adapted repository-specific data processing, splitting, training, metric computation, and logging to Eval4DiRec. 
Because these settings, datasets, and data-splitting protocols differ from those used in the original studies, the resulting scores are not directly comparable with the originally reported values, and numerical differences are expected. 
Such differences cannot be attributed solely to model implementation. The results are intended for controlled comparisons among models within Eval4DiRec rather than for direct numerical reproduction of the original studies. 

\begin{itemize}

\item \textbf{Collaborative Filtering:}
We implement four representative diffusion-based CF models: CODIGEM~\cite{walker2022recommendation}, DiffRec~\cite{wang2023diffusion}, CF-Diff~\cite{hou2024collaborative}, and DDRM~\cite{zhao2024denoising}, which incorporate diffusion mechanisms into traditional user-item interaction modeling.

\item \textbf{Sequential Recommendation:}
This category includes five models that incorporate diffusion processes into sequence modeling: 
DiffuRec~\cite{li2023diffurec}, DiffuASR~\cite{liu2023diffusion}, CaDiRec~\cite{cui2024context}, and CDDRec~\cite{wang2024conditional}. Additionally, T-DiffRec~\cite{wang2023diffusion}, which extends DiffRec with time-aware modeling, is included as a sequential recommendation variant.

\item \textbf{Multimodal Recommendation:}
For multimodal tasks, we support three diffusion-based models: MCDRec~\cite{ma2024multimodal}, DiffMM~\cite{jiang2024diffmm}, and CCDRec~\cite{yang2024curricular}, which leverage multimodal diffusion to enhance recommendation quality.

\item \textbf{POI Recommendation:}
We include Diff-POI~\cite{Qin2023diffpoi}, a diffusion-based recommendation model tailored for spatial and temporal modeling in point-of-interest recommendation.

\item \textbf{Cross-Domain Recommendation:}
We implement DiffCDR~\cite{xuan2024diffusion}, which employs a diffusion model to transfer knowledge across domains.

\end{itemize}

All supported models are implemented following a consistent architectural interface, making it straightforward to reproduce experiments within Eval4DiRec, conduct fair comparisons, and extend to new methods.

\subsection{Execution Module}

The execution module is designed to manage the training and evaluation processes of diffusion-based recommendation models in a unified and systematic manner. 
It coordinates the end-to-end workflow, including model training on the training split, model selection on the validation split, and final reporting on the test split, while keeping the protocol consistent across models. 
To support the five recommendation scenarios covered in Eval4DiRec, the module provides dedicated execution routines for each scenario. These routines are aligned with the corresponding data format and supervision signals, and the module also exposes configurable options for optimization and sampling during evaluation.
Through these features, the execution module enables comprehensive and reproducible benchmarking of various diffusion-based recommendation methods within a unified experimental environment, thereby promoting research and facilitating fair comparisons across models and tasks in the field.

More specifically, the execution process follows a unified train--validation--test protocol. 
Given a grouped YAML configuration, the Execution module first initializes the experimental context, including the random seed, computing device, dataset setting, model setting, diffusion configuration, optimization hyperparameters, and evaluation metrics. It then constructs the corresponding data loaders, recommendation model, diffusion process, optimizer, trainer, and evaluator according to the specified recommendation scenario. During training, the model is optimized on the training split under the configured recommendation and diffusion objectives, while its performance is periodically monitored on the validation split. The best checkpoint is selected according to a predefined validation criterion, such as Recall or NDCG, and is subsequently used for final evaluation on the test split. The module reports Top-$K$ recommendation metrics under the same evaluation protocol and records configurations, checkpoints, logs, metadata, and final results for reproducibility. This design standardizes the experimental lifecycle while still allowing model-specific optimization and sampling settings to be specified through configuration files.

\subsection{Utils Module}

The Execution module is further supported by the \textit{Utils} module, which provides shared utilities for the entire pipeline. These utilities include configuration parsing, component registration, random seed and device management, logging, and metric computation. By centralizing these common functions, Eval4DiRec prevents different models from silently using inconsistent metric definitions, seed initialization rules, logging formats, or configuration conventions. New datasets, models, trainers, evaluators, and metrics can be connected through the registry mechanism, which improves extensibility while preserving a common evaluation protocol.

\subsubsection{Metrics}

To ensure fair and comprehensive comparisons among models, we adopt four widely used Top-$K$ ranking metrics, including Recall, Normalized Discounted Cumulative Gain (NDCG), Hit Ratio (HR), and Mean Reciprocal Rank (MRR)~\cite{ma2024plug, zhu2024graph}. 
We report all metrics with $K{=}20$ in our experiments. 
Recall@K is defined as the fraction of ground-truth relevant items that appear in the Top-$K$ recommendation list. 
It primarily measures retrieval effectiveness, reflecting how well a model can cover relevant items within a limited recommendation list. 
NDCG@K is defined as a position-aware ranking metric that discounts gains at lower ranks, assigning higher weights to relevant items placed earlier in the list. 
It measures ranking quality by accounting for both relevance and position, and the normalization makes scores comparable across users. 
HR@K is defined as an indicator of whether at least one relevant item is retrieved within the Top-$K$ list for a user. 
It measures hit-based success, capturing whether the model can provide at least one correct recommendation under the Top-$K$ cutoff. 
MRR@K is defined as the reciprocal rank of the first relevant item in the Top-$K$ list, averaged over users. 
It measures how early the first correct recommendation appears, and is therefore sensitive to the topmost ranks in the recommendation list.

\subsection{Extensibility to Custom Models}

One of the key design goals of Eval4DiRec is to allow researchers to incorporate newly developed diffusion-based recommender models without changing the established evaluation protocol. 
To achieve this, Eval4DiRec adopts a modular architecture that decouples model implementation from data processing, training routines, and evaluation workflows. 
As a result, integrating a new model mainly requires implementing its core modeling logic, without repeatedly writing common components such as data reading, training loops, or metric computation. 
More importantly, this decoupling helps keep data inputs, training procedures, and evaluation implementations consistent across models, reducing discrepancies introduced by implementation details.

Eval4DiRec achieves extensibility through configuration-driven component composition. 
A grouped YAML configuration specifies the dataset, recommendation scenario, model, diffusion process, optimizer, evaluation metrics, and output settings. 
At runtime, the registry maps these entries to the corresponding data loaders, scenario adapters, recommenders, diffusion processes, trainers, and evaluators. 
This design decouples experiment specification from component implementation, allowing new methods to reuse the existing data processing, training, evaluation, and logging pipeline while extending only the components with method-specific functionality.

For a new model that follows an existing recommendation scenario, the developer typically implements only a recommender class under \texttt{eval4direc/models/} and provides a YAML configuration under \texttt{configs/}. 
The recommender inherits the scenario-aware base interface and implements \texttt{calculate\_loss} and \texttt{predict}. 
The \texttt{calculate\_loss} interface consumes standardized batches from the Data module and returns the training objective, while \texttt{predict} produces user-item scores or ranking signals for unified Top-$K$ evaluation. 
The data splitting, batching, training loop, validation-based model selection, checkpoint management, metric computation, and result logging are reused from the framework. 
Therefore, developers do not need to create method-specific dataset classes, trainers, or evaluators unless the original algorithm introduces a real behavioral difference.

When method-specific differences are necessary, Eval4DiRec supports localized extension through registries. 
For example, a model with special sample construction can register a data profile while reusing the scenario-level loader; a method with a customized denoising procedure can register its diffusion process, sampler, schedule, or noise distribution without changing the recommendation backbone; and a new task protocol can be introduced through a scenario adapter while retaining the common trainer and evaluator. 
This mechanism confines method-specific changes to the relevant components, thereby reducing boilerplate code and avoiding unnecessary modifications to data processing, training logic, configuration parsing, and evaluation scripts.

Eval4DiRec adopts a modular architecture to decouple the major stages of diffusion-based recommendation evaluation. Under \texttt{eval4direc/}, runtime contracts, data abstractions, scenario adapters, recommender models, diffusion processes, samplers, trainers, and evaluators are implemented as separate modules with registry-based component instantiation. This design provides a unified interface for heterogeneous diffusion-based recommenders, while allowing scenario-specific data processing, model training, reverse sampling, and metric computation to be configured independently. As a result, new models and scenarios can be incorporated without changing the shared evaluation protocol.

As a concrete example, we integrated T-DiffRec into Eval4DiRec as a time-aware sequential recommendation model without introducing a model-specific trainer, evaluator, metric implementation, or experiment entry point. 
The recommender is registered under \texttt{eval4direc/models/sequential/t-\\diffrec.py} and implements the same \texttt{calculate\_loss} and \texttt{predict} interfaces used by the shared execution pipeline. Because T-DiffRec differs from standard sequential models in its construction of time-weighted interaction vectors, this behavior is isolated in a model-specific data profile under \texttt{eval4direc/data/profiles/sequential/tdiffrec.py}. The profile reuses the common sequential data loader and leave-one-out protocol, while assigning weights to each user's chronologically ordered training interactions. Dataset-specific settings, including the model name, data path, diffusion process, optimization parameters, and evaluation metrics, are declared in grouped YAML files under \texttt{configs/tdiffrec/}. All remaining functionality, including data loading, training orchestration, validation-based checkpoint selection, Top-$K$ evaluation, metric computation, and result logging, is inherited from the shared framework. This example demonstrates that a method-specific extension can be confined to the recommender, the data profile required by its distinctive input construction, and the corresponding configurations, without modifying the common training and evaluation infrastructure.

\section{Experiment Evaluation}
\label{sec:Evaluation}

The experiments are designed to evaluate not only the performance of diffusion-based recommender systems, but also the effectiveness of Eval4DiRec as a unified evaluation framework. Specifically, we examine whether Eval4DiRec can support fair comparison under consistent data processing, splitting, training, inference, and metric protocols; manage diffusion-specific configurations such as noise schedules, diffusion steps, noise distributions, and sampling settings; and provide diagnostic analyses on efficiency, robustness, and long-tail recommendation. 

In this section, to comprehensively explore the potential of diffusion-based recommender systems, we conduct a systematic and thorough evaluation aimed at addressing the following research questions: \\
\textbf{RQ1}: How is the overall performance of diffusion-based RSs? \\
\textbf{RQ2}: What are the key factors influencing the performance of diffusion-based RSs? \\
\textbf{RQ3}: How robust are diffusion-based RSs when exposed to noisy data and missing interactions? \\
\textbf{RQ4}: Can diffusion-based RSs alleviate the popularity bias and improve the recommendation performance for long-tail items?

\subsection{Experiment Settings}
\subsubsection{Baselines}

To ensure a rigorous performance benchmark, we incorporate both state-of-the-art diffusion-based models and representative non-diffusion baselines for each scenario.

\textbf{Collaborative Filtering}:
\textbf{BPR-MF}~\cite{rendle2009bpr} is a classical collaborative filtering method that learns user and item embeddings through matrix factorization.
\textbf{LightGCN}~\cite{he2020lightgcn} is a graph-based method that learns representations via linear neighborhood aggregation on the user–item interaction graph.
\textbf{CODIGEM}~\cite{walker2022recommendation} models users' historical interactions through the forward process and generates item ratings in the reverse process to achieve ranking prediction. 
\textbf{DiffRec}~\cite{wang2023diffusion} takes binary user interaction vectors as input, optimizes performance by reducing the noise scale and decreasing the number of forward diffusion steps, and directly predicts the target from interaction history.
\textbf{CF-Diff}~\cite{hou2024collaborative} incorporates multi-hop neighbor information and cross-attention mechanisms during the reverse denoising process to deeply mine high-order collaborative signals. 
\textbf{DDRM}~\cite{zhao2024denoising} enhances the robustness of user and item embeddings through multi-step denoising, starting with the average embedding of users' historical interactions and using collaborative information to guide the denoising process.

\textbf{Sequential}:
\textbf{BERT4Rec}~\cite{sun2019bert4rec} improves sequential recommendation by using a bidirectional Transformer with a cloze task, addressing the limitations of uni-directional models like SASRec.
\textbf{STOSA}~\cite{fan2022sequential} enhances self-attention with a Wasserstein-based mechanism to capture pairwise item correlations and introduces uncertainty through stochastic Gaussian modeling.
\textbf{DiffuRec}~\cite{li2023diffurec} models target items as probability distributions and leverages diffusion models to generate diverse item representations. 
\textbf{DiffuASR}~\cite{liu2023diffusion} generates high-quality pseudo-sequences based on historical sequences, enhancing generation performance through a sequential U-Net architecture and guidance strategies. 
\textbf{T-DiffRec~\cite{wang2023diffusion} extends DiffRec with time-aware interaction modeling for sequential recommendation.}
\textbf{CaDiRec}~\cite{cui2024context} proposes a context-aware diffusion contrastive learning method, fusing contextual information to create augmented views, thereby improving the model's representation capabilities. 
\textbf{CDDRec}~\cite{wang2024conditional} designs a conditional denoising diffusion model that uses historical sequences to guide the denoising process and incorporates cross-attention mechanisms to generate high-quality representations.

\textbf{Multimodal}:
\textbf{BM3}~\cite{zhou2023bootstrap} employs a multi-modal contrastive learning strategy, incorporating inter- and intra-modality contrastive losses to strengthen representation learning.
\textbf{MCDRec}~\cite{ma2024multimodal} integrates modality features to generate item representations and denoises the user-item graph using a diffusion model. 
\textbf{DiffMM}~\cite{jiang2024diffmm} leverages a diffusion model to generate multimodal-aware user-item graphs and optimizes user representations through cross-modal contrastive learning.
\textbf{CCDRec}~\cite{yang2024curricular} employs a diffusion model to align multimodal features with collaborative signals and progressively generates adaptive negative samples for optimized training.

\textbf{Point of Interest}:
\textbf{STAN}~\cite{luo2021stan} captures sequential patterns via a spatio-temporal attention network, incorporating personalized item frequency for recommendation.
\textbf{Diff-POI}~\cite{Qin2023diffpoi} explores user visitation patterns through diffusion sampling and optimizes the scoring function to generate recommendations.

\textbf{Cross-domain}:
\textbf{PTUPCDR}~\cite{zhu2022personalized} is a meta-learning-based cross-domain recommendation method that generates user-specific mapping functions to transfer knowledge across domains.
\textbf{DiffCDR}~\cite{xuan2024diffusion} uses a diffusion model to generate user embeddings in the target domain, conditioned on the source domain embeddings. It optimizes cross-domain recommendation performance through similarity alignment.

\subsubsection{Implementation Details}

Our experiments are conducted on a deep learning server running Ubuntu 22.04, equipped with an NVIDIA RTX 3090 GPU, Python 3.10, and PyTorch 2.6. 
To ensure fair comparisons, we tune hyperparameters for both diffusion-based models and non-diffusion baselines according to validation performance, and run all methods under the same data processing, training, and evaluation pipeline implemented in our unified framework. 
We start from the hyperparameter settings reported in the original papers and further refine them within a predefined search space to obtain the best validation performance across datasets~\cite{wang2024conditional, jiang2024diffmm}. 
We employ AdamW for optimization. The learning rate is selected from $\{1\mathrm{e}{-5}, 5\mathrm{e}{-5}, 1\mathrm{e}{-4}, 5\mathrm{e}{-4}, 1\mathrm{e}{-3}\}$, and the L2 regularization coefficient is selected from $\{0, 1\mathrm{e}{-5}, 1\mathrm{e}{-4}\}$. 
The hidden dimension is tuned from $\{64, 128, 256\}$. For sequential models, the maximum user history length is chosen from $\{10, 20, 30, 50, 100, 200\}$ depending on the dataset. 
For diffusion-based models, we examine multiple noise schedules, including \textit{linear}, \textit{exponential}, \textit{sqrt}, and \textit{cosine}. 
The number of diffusion steps is selected from $\{2, 3, 5, 10, 20, 50, 100, 200, 500, 1000\}$, depending on the design choices of each model. 
Throughout the diffusion process, we use a standard Gaussian noise distribution. 
We train each model for up to 1000 epochs with early stopping on the validation set, stopping if the validation performance does not improve for 20 consecutive checks. 
Each experiment is repeated with five different random seeds, and we report the average performance.

\begin{table}[ht]
    \renewcommand{\arraystretch}{0.89}
    \setlength{\tabcolsep}{1pt} 
    \small
    \centering
    \caption{Overall Performance Comparison. Different recommendation tasks are evaluated on their respective datasets. All metrics are reported @20 and shown as percentages (\%). Bold and underlined values indicate the best and second-best results on each dataset, respectively. Results marked with * are statistically significant based on a paired $t$-test ($p<0.05$). Models marked with \dag\ are non-diffusion baselines.}
    \begin{tabular}{l|cccc|cccc|cccc}
        \hline
        {\textbf{Method}} &{\textbf{Recall}}&{\textbf{NDCG}}&{\textbf{MRR}}&{\textbf{HR}} &{\textbf{Recall}}&{\textbf{NDCG}}&{\textbf{MRR}}&{\textbf{HR}} &{\textbf{Recall}}&{\textbf{NDCG}}&{\textbf{MRR}}&{\textbf{HR}}\\
        \hline \hline
        \textit{\textbf{CF}}    & \multicolumn{4}{c|}{\textbf{Yelp}} & \multicolumn{4}{c|}{\textbf{Beauty}} & \multicolumn{4}{c}{\textbf{ML-1M}} \\
        \hline
        \textbf{BPR-MF\dag}& 1.81      & 1.03    & 1.12& 6.71  & 4.49      & 2.73  & 2.84& 9.43 & 9.75  & 6.63  & 11.57&42.16 \\
        \textbf{LightGCN\dag}& 3.31& 1.79& 2.14& 9.09 & 6.32& 3.55& 3.61& 11.86 & 12.31& 8.74& 14.06&48.28 \\
        \textbf{CODIGEM}&\underline{4.12}      &\underline{2.24}    &  \underline{2.61}&  \underline{10.93}  &6.90& 4.02&\underline{4.09}&  13.67&\underline{15.56}&\underline{10.97}& 17.07&57.46\\
        \textbf{DiffRec}& \textbf{4.82*}&\textbf{2.66*}&  \textbf{3.16*}&  \textbf{12.69*}& 7.13&  \underline{4.08}&4.06&  14.04& \textbf{15.67*}&  \textbf{11.05*}& \textbf{18.49*}&\textbf{58.02*}\\
        \textbf{CF-Diff}& 3.36& 1.91&  2.34&  9.28& \underline{7.18}&  3.94&3.89  &  \underline{14.21}& 15.52& 10.76& \underline{17.62}&\underline{57.72}\\
        \textbf{DDRM}& 4.01& 2.13&  2.22&  10.21&  \textbf{7.60*}&  \textbf{4.21*}&\textbf{4.26*}&  \textbf{14.45*}& 15.03& 10.32& 15.71&54.72\\
        \hline
        \hline
        \textit{\textbf{Sequential}}    & \multicolumn{4}{c|}{\textbf{Yelp}} & \multicolumn{4}{c|}{\textbf{Beauty}} & \multicolumn{4}{c}{\textbf{ML-1M}}  \\
        \hline
        \textbf{BERT4Rec\dag}& 0.73& 0.42& 0.31& 0.73& 1.14& 0.72& 0.59& 1.14& 3.55& 1.31& 0.69&3.55\\
        \textbf{STOSA\dag}& 1.32& 0.75& 0.59& 1.32& 3.51& 1.54& 1.27& 3.51& 4.77& 2.03& 1.12&4.77\\
        \textbf{DiffuRec}&2.51&1.24&1.03&2.51&4.85&1.98&1.31&4.85&7.89&3.28&2.07&7.89\\
        \textbf{DiffuASR}& 0.94&  0.67&0.53&0.94& 1.47&  0.83&0.64&1.47& 3.79&1.43&0.75& 3.79\\
        \textbf{T-DiffRec}
        & 4.03&  1.65&0.98&4.03& 6.38&  2.07&\underline{1.65}&6.38& \underline{10.13}&\underline{4.16}&\underline{2.42}& \underline{10.13}\\
        \textbf{CaDiRec}& \underline{4.12}& \underline{1.97}&\underline{1.14}&\underline{4.12}&  \underline{6.41}& \underline{2.14}&{1.37}& \underline{6.41}&{9.16}&{4.13}&{2.39}&{9.16}\\
        \textbf{CDDRec}& \textbf{5.19*}& \textbf{2.13*} &\textbf{1.34*}& \textbf{5.19*} &  \textbf{7.29*} & \textbf{2.76*} &\textbf{1.71*}& \textbf{7.29*}  &\textbf{10.60*} &\textbf{4.51*} &\textbf{2.67*}&\textbf{10.60*}  \\
        \hline
        \hline
        \textit{\textbf{Multimodal}}    & \multicolumn{4}{c|}{\textbf{TikTok}} & \multicolumn{4}{c|}{\textbf{Baby}} & \multicolumn{4}{c}{\textbf{Sports}}  \\
        \hline
        \textbf{BM3\dag}& 7.31& 2.97& 2.01& 7.31 & 6.22& 2.65& 1.54& 6.46 & 4.48& 1.95& 1.28&4.98 \\
        \textbf{MCDRec}&  7.47  & 3.29&2.12&7.47   & 6.41  &  2.63 &1.64&6.41   & 4.92  & 2.07 &1.39&5.23 \\
        \textbf{DiffMM}&  \textbf{10.99*} & \textbf{4.72*}&\textbf{3.01*}&\textbf{10.99*}   & \underline{9.14} &  \underline{3.81} &\underline{2.41}&\underline{9.51}  & \underline{9.02}  & \underline{4.03} &\underline{2.69}&\underline{9.47}  \\
        \textbf{CCDRec}& \underline{9.41}& \underline{3.77}& \underline{2.25}& \underline{9.41}& \textbf{9.22*}& \textbf{3.93*}& \textbf{2.53*}& \textbf{9.59*}& \textbf{9.49*}& \textbf{4.14*}& \textbf{2.77*}& \textbf{9.93*}\\
        \hline
        \hline
        \textit{\textbf{POI}}    & \multicolumn{4}{c|}{\textbf{TKY}} & \multicolumn{4}{c|}{\textbf{NYC}} & \multicolumn{4}{c}{\textbf{Gowalla}}  \\
        \hline
        \textbf{STAN\dag}& \underline{67.51}& \underline{58.76}&\underline{54.92}&\underline{67.51} & \underline{64.17}& \underline{60.94}&\underline{59.16}&\underline{64.17} &  \underline{30.17}&  \underline{22.46}&  \underline{20.74}& \underline{30.17} \\
        \textbf{Diff-POI}& \textbf{70.10*} & \textbf{62.76*} &\textbf{60.41*}&\textbf{70.10*}  & \textbf{66.44*} & \textbf{62.61*} &\textbf{61.51*}&\textbf{66.44*}  &  \textbf{35.21*} &  \textbf{26.79*} &  \textbf{24.37*} & \textbf{35.21*}   \\
        \hline
        \hline
        \textit{\textbf{Cross-Domain}}    & \multicolumn{4}{c|}{\textbf{Video - Music}} & \multicolumn{4}{c|}{\textbf{Book - Video}} & \multicolumn{4}{c}{\textbf{Book - Music}}  \\
        \hline
        \textbf{PTUPCDR\dag}& \underline{1.46}& \underline{1.54}&\underline{4.13}&\underline{18.35} &  \underline{1.74}&  \underline{1.28}&\underline{2.76}&\underline{14.15} &  \underline{1.69}&  \underline{1.71}&\underline{3.12}&\underline{16.98} \\
        \textbf{DiffCDR}& \textbf{1.88*} & \textbf{1.94*} &\textbf{4.54*}&\textbf{19.01*}  & \textbf{1.86*}&  \textbf{1.44*} &\textbf{2.98*}&\textbf{15.09*}  &  \textbf{1.88*} & \textbf{1.75*} &\textbf{3.55*}&\textbf{17.55*}  \\
        \hline
    \end{tabular}
    \label{tab:overall}
\end{table}

\subsection{Overall Performance Comparison (RQ1)}

This experiment is designed to validate whether Eval4DiRec can support fair and comparable evaluation across heterogeneous diffusion-based recommender models. 
All methods are evaluated under the same data preprocessing, splitting strategy, metric implementation, and Top-$K$ evaluation protocol within each recommendation scenario. 
The overall performance is shown in Table~\ref{tab:overall}. We report the experimental results of Recall, NDCG, MRR, and HR.

In collaborative filtering, diffusion-based methods generally outperform the non-diffusion baselines, with DiffRec and DDRM showing the strongest results across the three datasets. 
Specifically, DiffRec achieves the best performance on Yelp and ML-1M, while DDRM performs best on Beauty. 
This trend may be attributed to the fact that  both methods cast collaborative filtering as a progressive denoising generation process, which can more reliably recover user preference signals from sparse or noisy interactions. 
Moreover, they adopt inference designs that are better aligned with recommendation than standard diffusion setups, such as avoiding fully corrupting signals into pure noise or injecting collaborative information as guidance during denoising, which helps preserve personalization and improves ranking quality. 
In contrast, CODIGEM and CF-Diff remain competitive but exhibit less consistent advantages across datasets, suggesting that the configuration of the sampling process and the inference procedure can substantially influence performance in collaborative filtering.

In the sequential recommendation setting, CDDRec achieves the best performance on all three datasets, while CaDiRec and T-DiffRec are also highly competitive. 
T-DiffRec outperforms DiffuRec and DiffuASR on most metrics and obtains the second-best performance on ML-1M, suggesting that time-aware interaction information can help diffusion-based models better capture temporal preference patterns. 
This pattern aligns with the fact that sequential recommendation is particularly sensitive to how conditional information is incorporated, as stronger methods typically leverage sequential context to constrain the denoising process, making the sampled signals more consistent with users’ recent interests and behavioral patterns. 
In addition, approaches that employ context-aware diffusion augmentation or combine diffusion with contrastive learning can improve the discriminability and robustness of sequence representations, leading to more stable gains on Top-$K$ ranking metrics. 
By contrast, if the denoising process is not sufficiently guided by sequence context, the model may underutilize long-term signals or fail to adapt to preference changes, leading to worse ranking performance.

In the multimodal recommendation setting, DiffMM achieves the best performance on TikTok, while CCDRec performs best on Baby and Sports, and both methods outperform the non-diffusion baseline as well as the other diffusion models. 
This outcome is consistent with the fact that multimodal recommendation largely hinges on how multimodal features are aligned with collaborative signals. 
Stronger methods tend to explicitly capture cross-modal correlations during diffusion modeling and integrate multimodal semantics with interaction structure through denoising objectives or contrastive learning.
In addition, CCDRec further strengthens training by introducing more effective negative sampling strategies or difficulty scheduling, which helps sharpen the decision boundary and yields more stable ranking results. 
Overall, these results suggest that diffusion can contribute not only through generation or reconstruction, but also by serving as a training mechanism that enhances multimodal representation learning.

\begin{figure*}[t!]
    \centering
    \begin{subfigure}[b]{0.32\textwidth}
        \includegraphics[width=\linewidth]{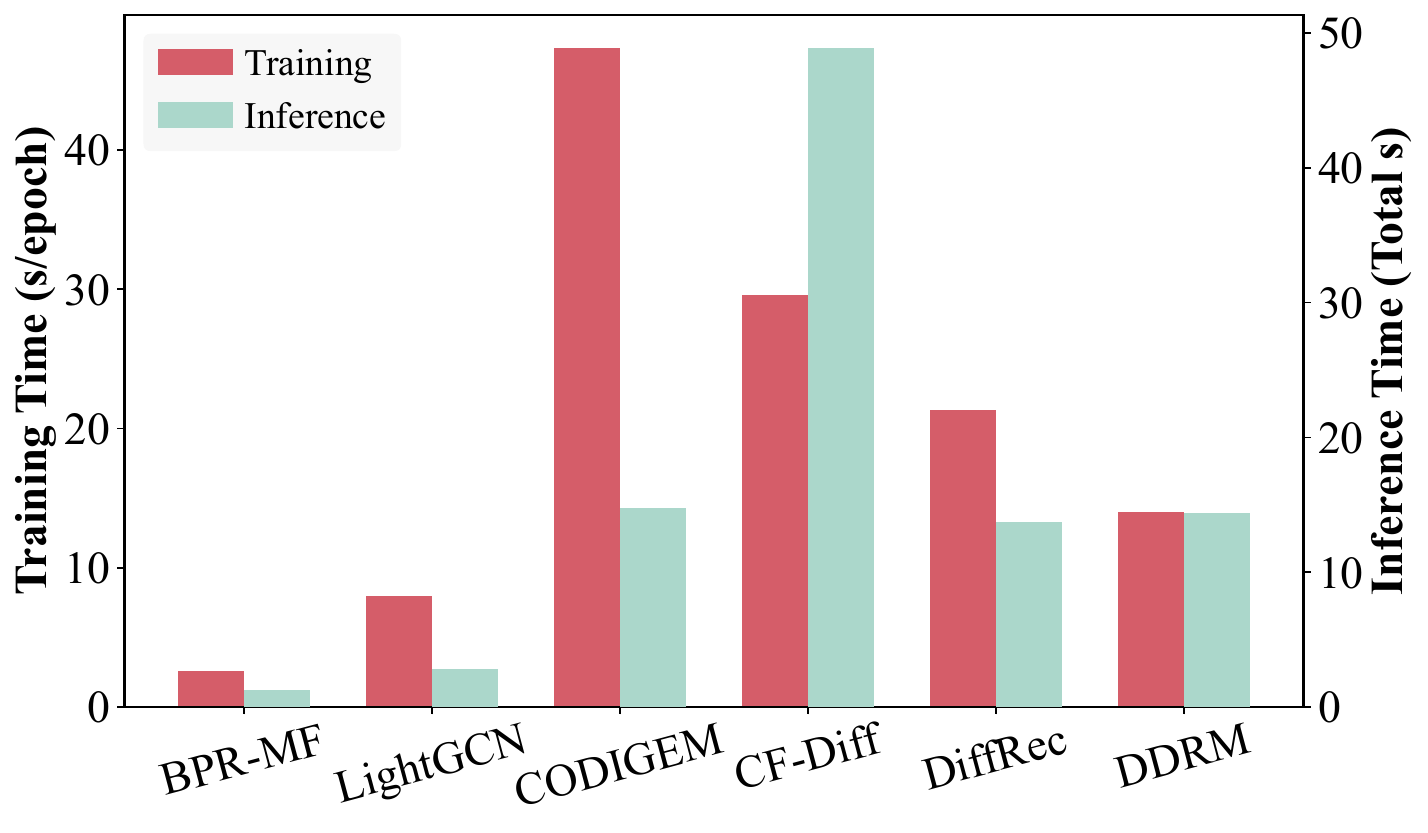}
        \captionsetup{labelfont=normalfont, textfont=normalfont}
        \caption{CF - Yelp}
        \label{fig:eff-cf-yelp}
    \end{subfigure}
    \begin{subfigure}[b]{0.32\textwidth}
        \includegraphics[width=\linewidth]{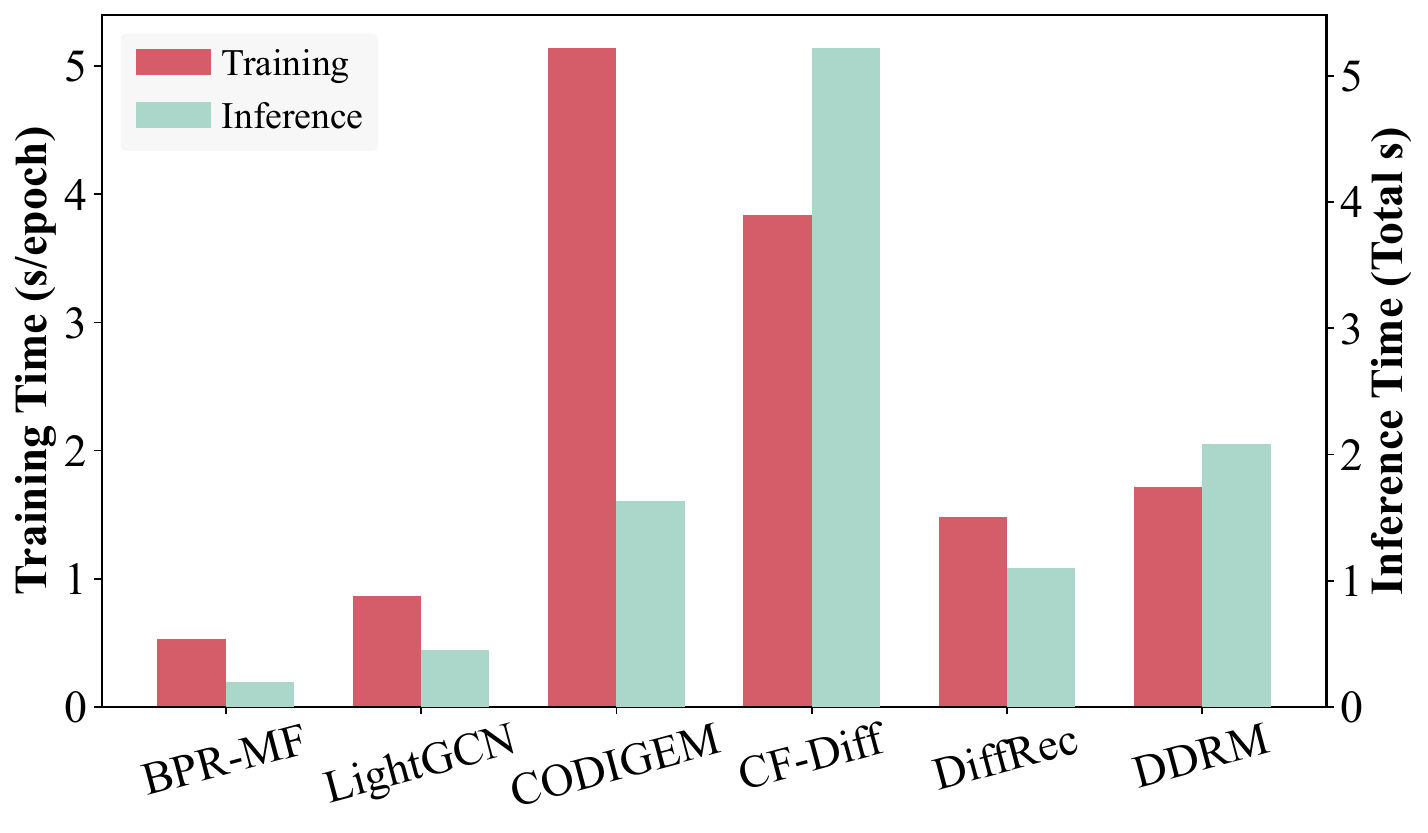}
        \captionsetup{labelfont=normalfont, textfont=normalfont}
        \caption{CF - Beauty}
        \label{fig:eff-cf-beauty}
    \end{subfigure}
    \begin{subfigure}[b]{0.32\textwidth}
        \includegraphics[width=\linewidth]{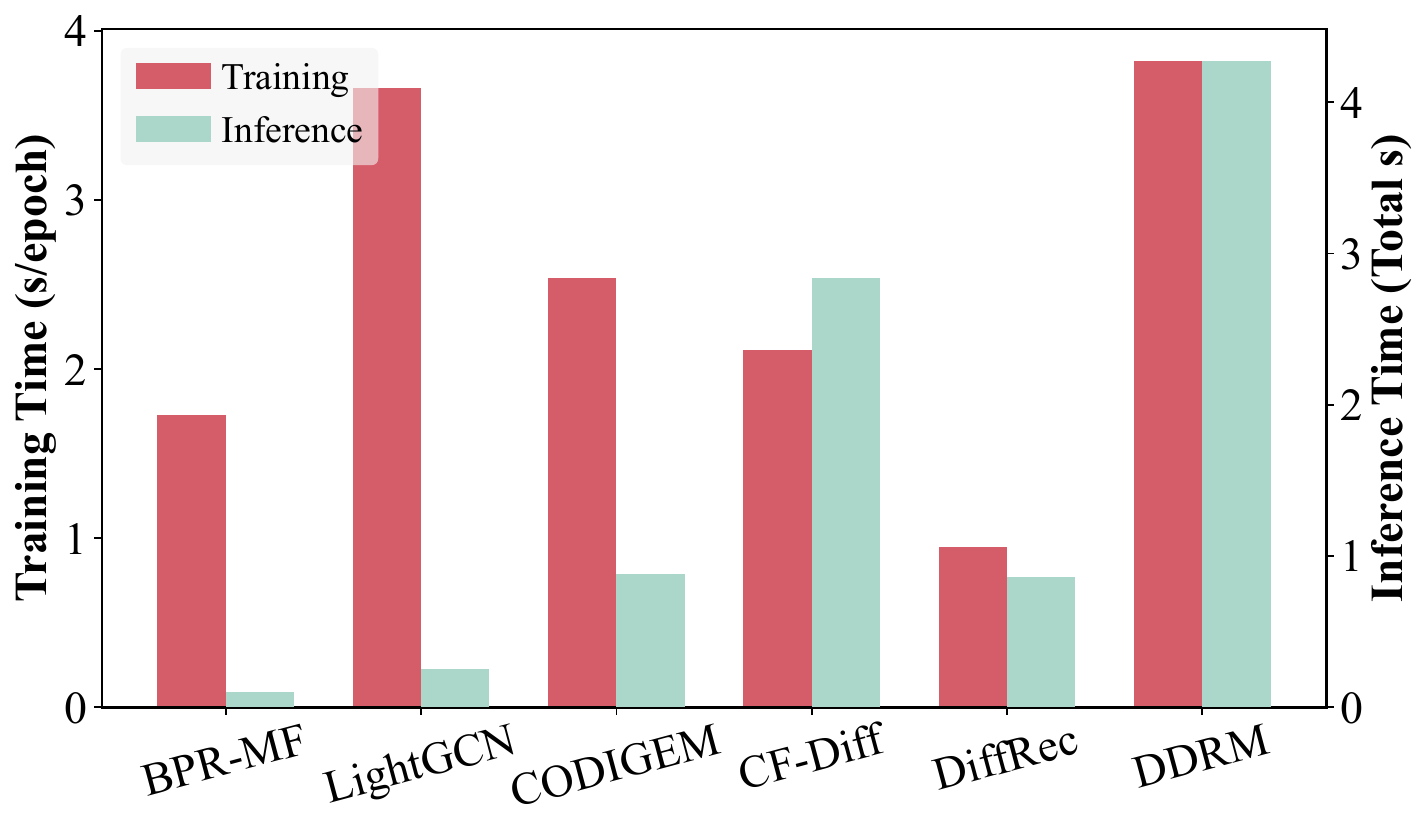}
        \captionsetup{labelfont=normalfont, textfont=normalfont}
        \caption{CF - ML-1M}
        \label{fig:eff-cf-ml1m}
    \end{subfigure}

    \begin{subfigure}[b]{0.32\textwidth}
        \includegraphics[width=\linewidth]{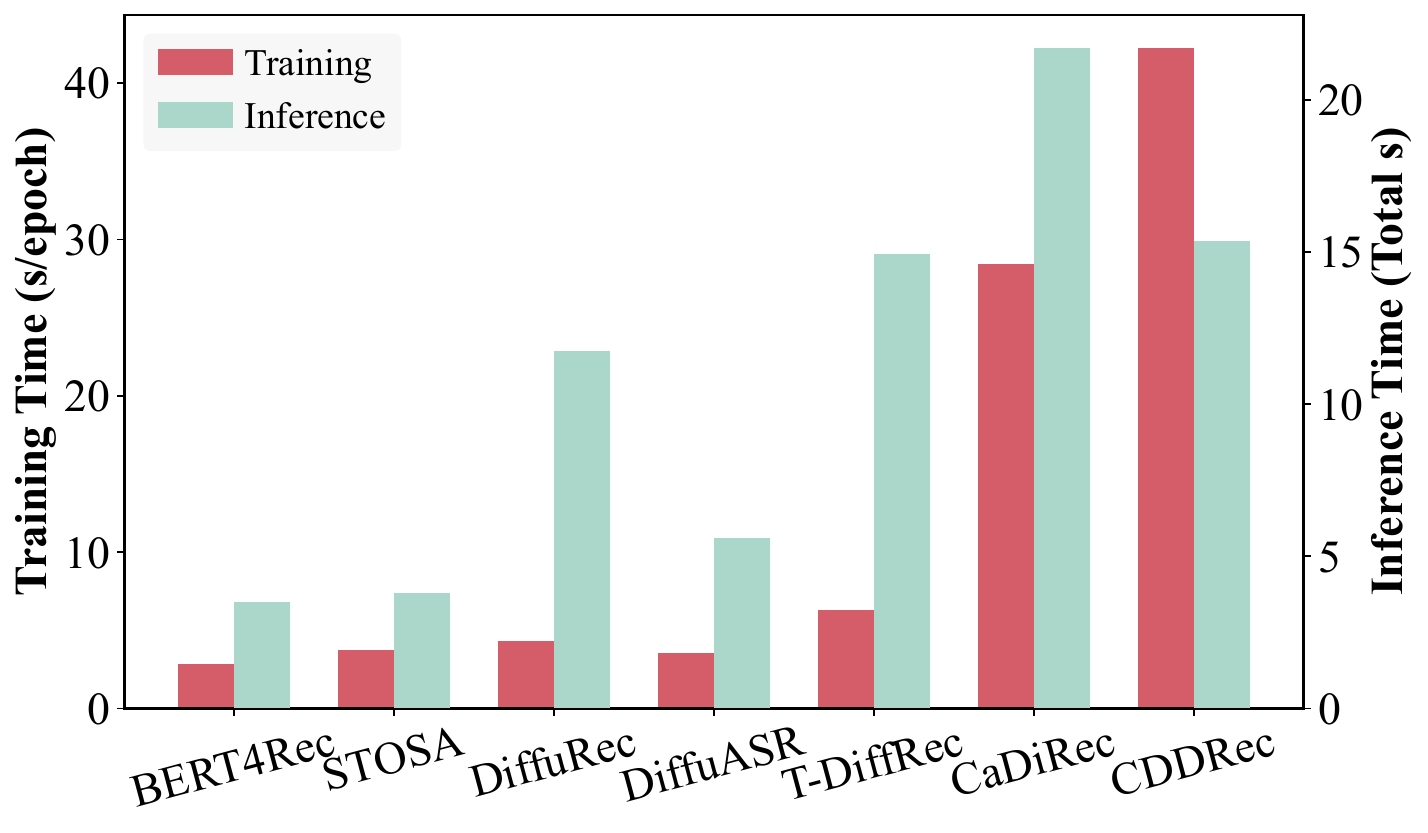}
        \captionsetup{labelfont=normalfont, textfont=normalfont}
        \caption{Sequential - Yelp}
        \label{fig:eff-seq-yelp}
    \end{subfigure}
    \begin{subfigure}[b]{0.32\textwidth}
        \includegraphics[width=\linewidth]{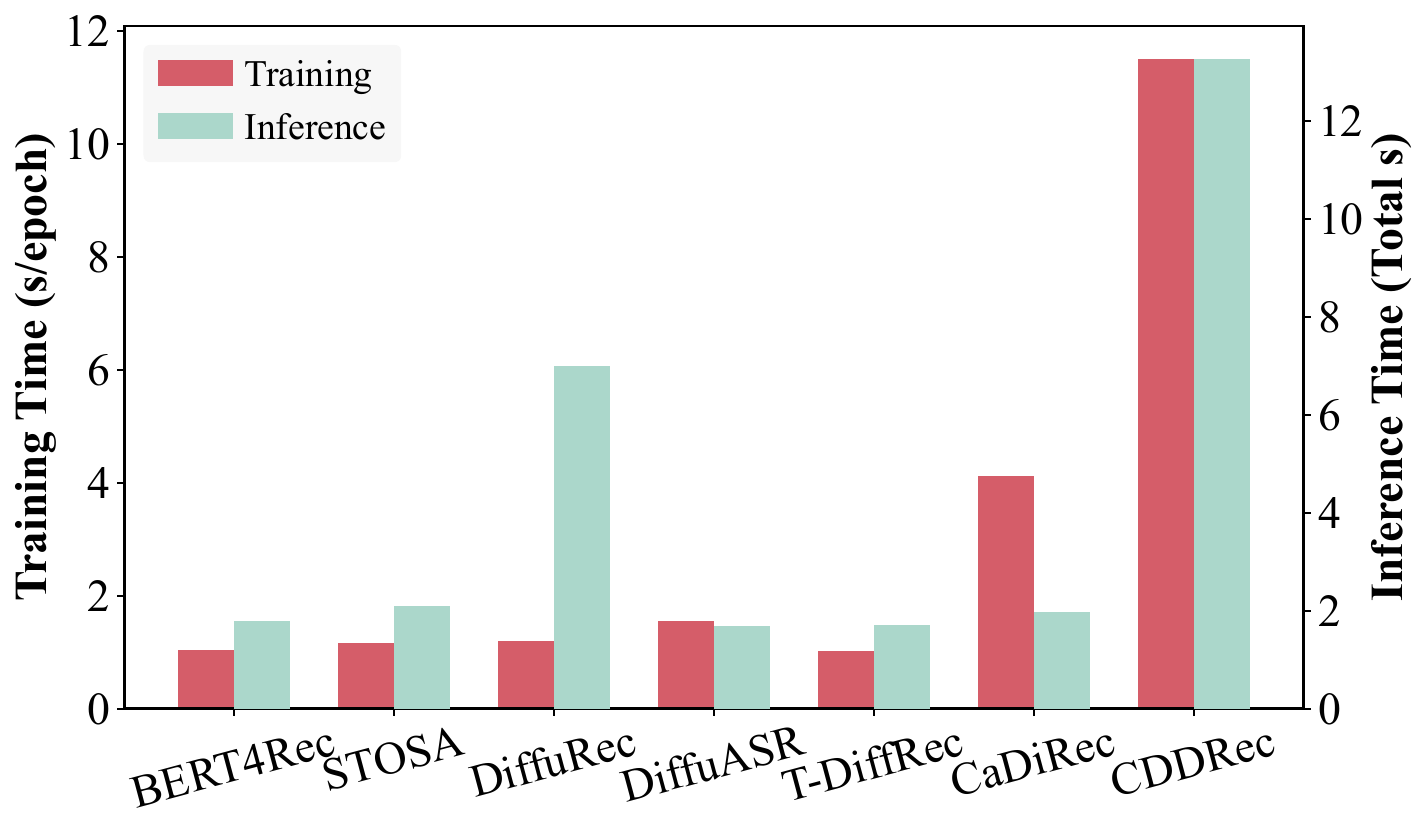}
        \captionsetup{labelfont=normalfont, textfont=normalfont}
        \caption{Sequential - Beauty}
        \label{fig:eff-seq-beauty}
    \end{subfigure}
    \begin{subfigure}[b]{0.32\textwidth}
        \includegraphics[width=\linewidth]{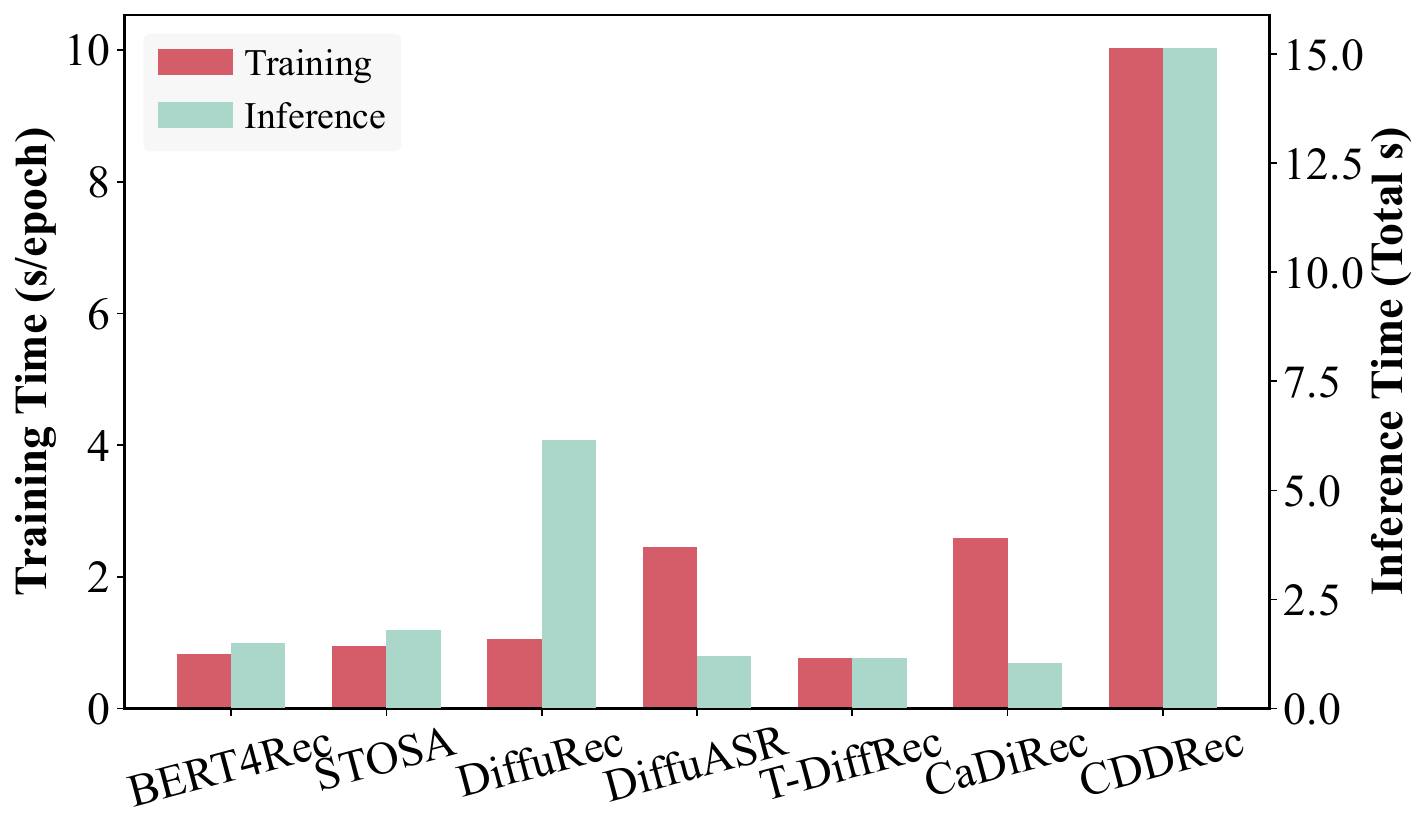}
        \captionsetup{labelfont=normalfont, textfont=normalfont}
        \caption{Sequential - ML-1M}
        \label{fig:eff-seq-ml1m}
    \end{subfigure}

    \begin{subfigure}[b]{0.32\textwidth}
        \includegraphics[width=\linewidth]{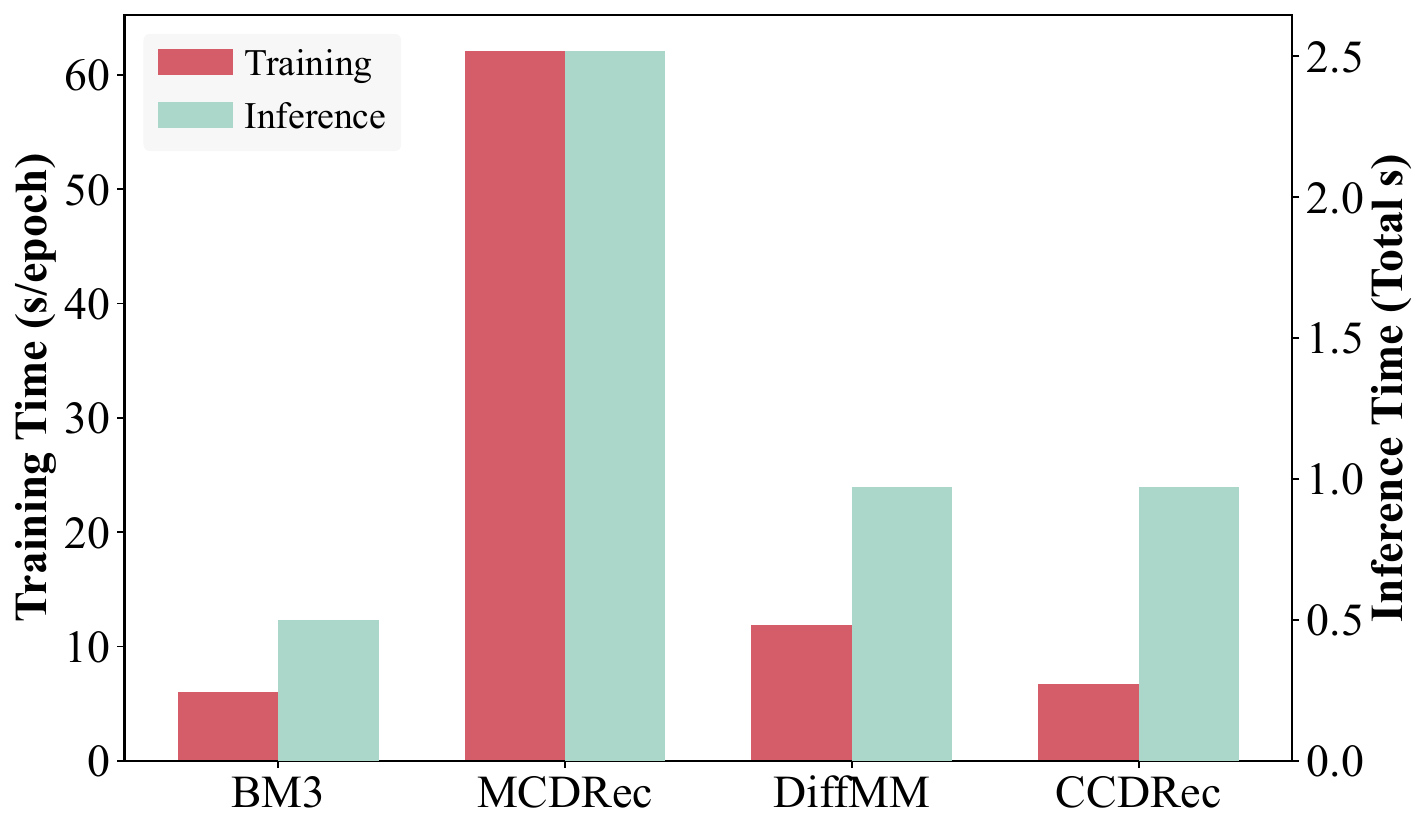}
        \captionsetup{labelfont=normalfont, textfont=normalfont}
        \caption{Multimodal - TikTok}
        \label{fig:eff-mm-tiktok}
    \end{subfigure}
    \begin{subfigure}[b]{0.32\textwidth}
        \includegraphics[width=\linewidth]{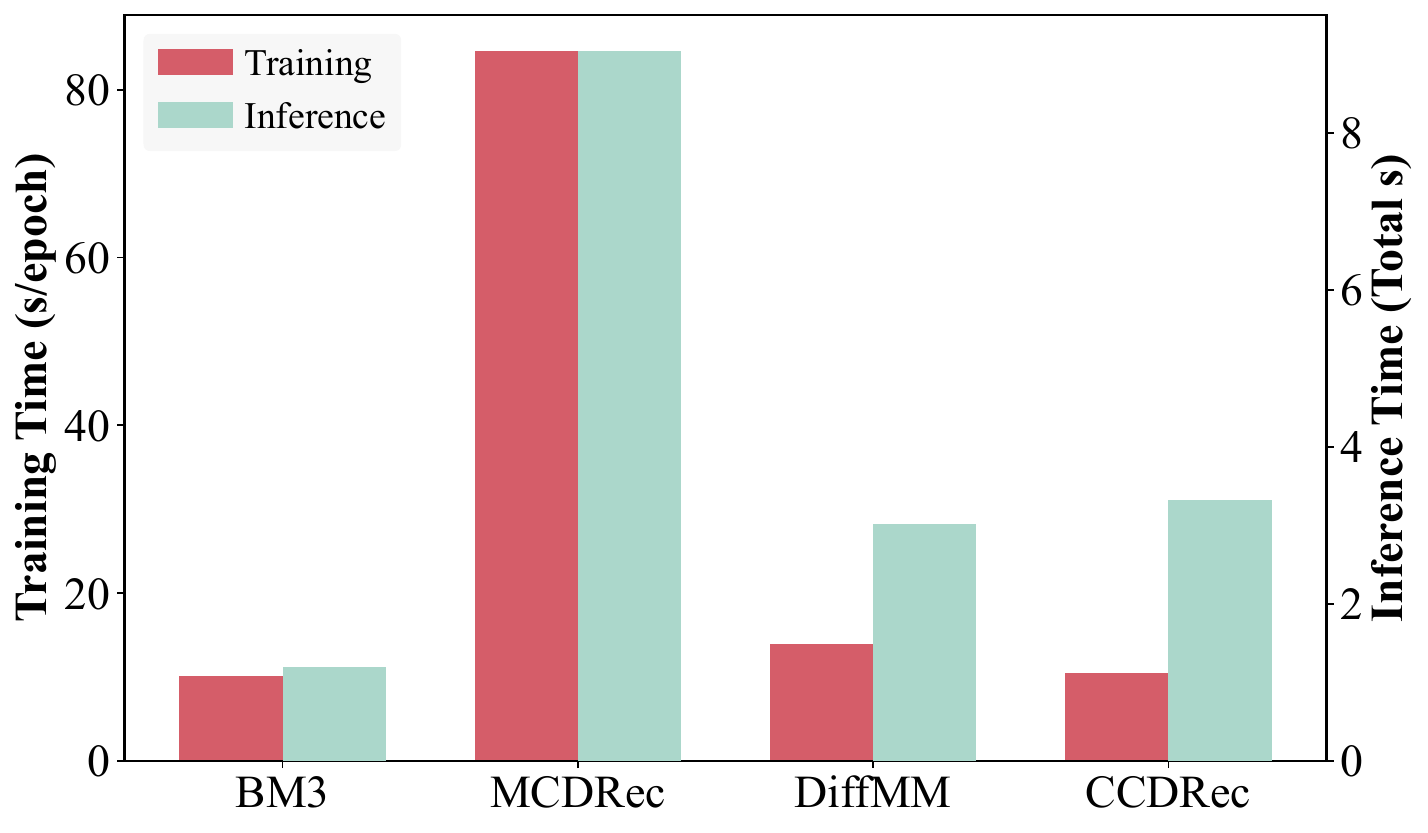}
        \captionsetup{labelfont=normalfont, textfont=normalfont}
        \caption{Multimodal - Baby}
        \label{fig:eff-mm-baby}
    \end{subfigure}
    \begin{subfigure}[b]{0.32\textwidth}
        \includegraphics[width=\linewidth]{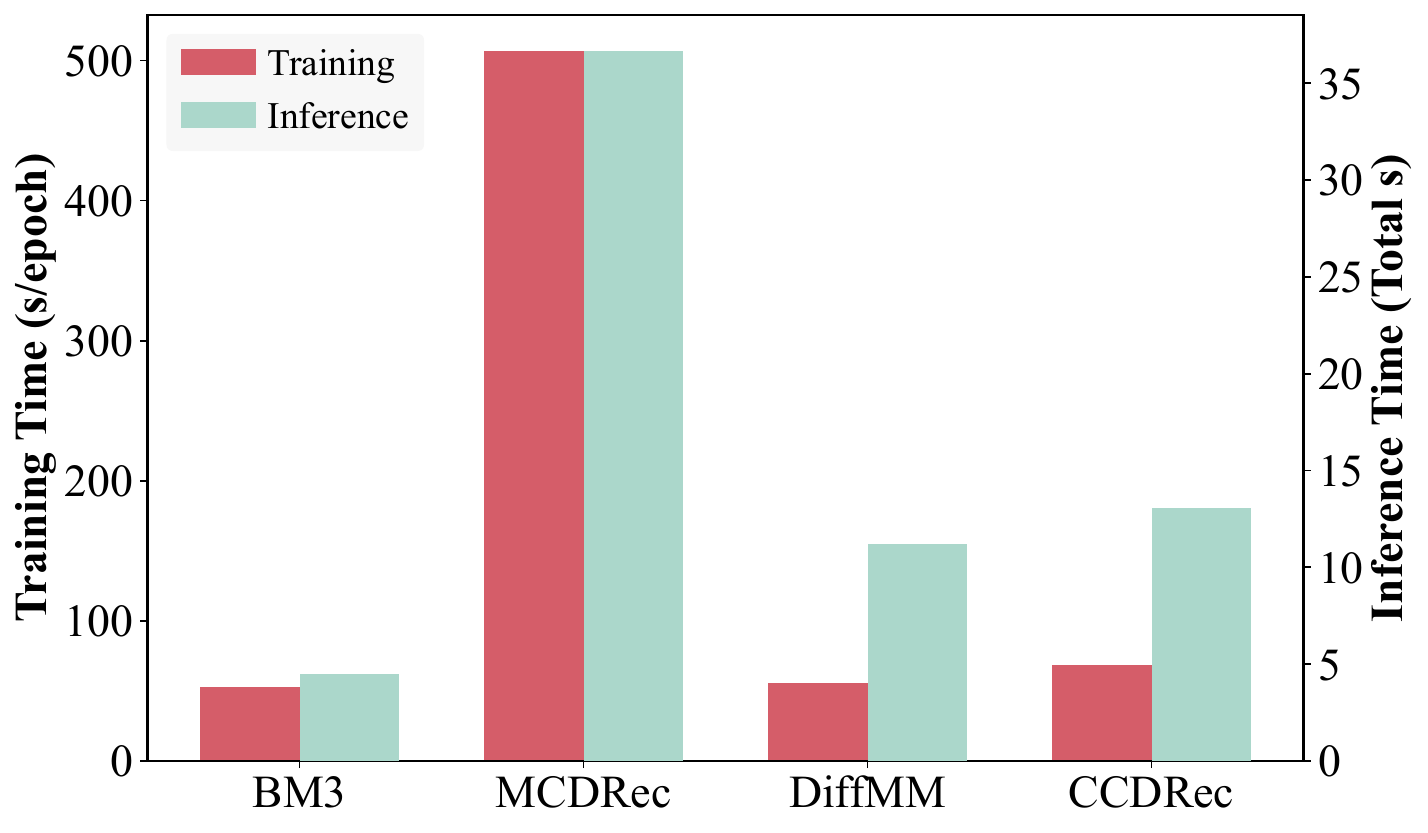}
        \captionsetup{labelfont=normalfont, textfont=normalfont}
        \caption{Multimodal - Sports}
        \label{fig:eff-mm-sports}
    \end{subfigure}

    \begin{subfigure}[b]{0.48\textwidth}
        \includegraphics[width=\linewidth]{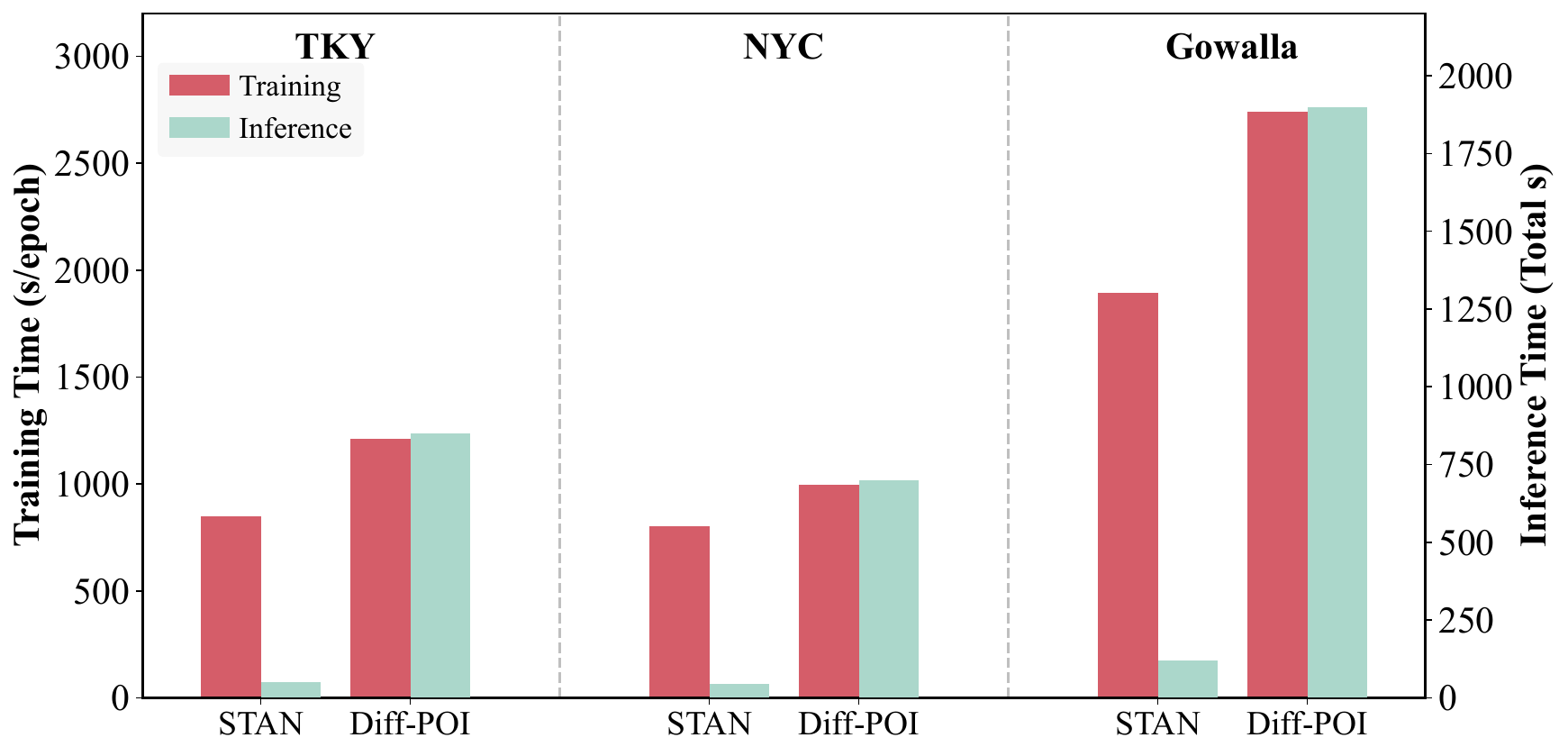}
        \captionsetup{labelfont=normalfont, textfont=normalfont}
        \caption{POI}
        \label{fig:eff-poi-merged}
    \end{subfigure}
    \begin{subfigure}[b]{0.48\textwidth}
        \includegraphics[width=\linewidth]{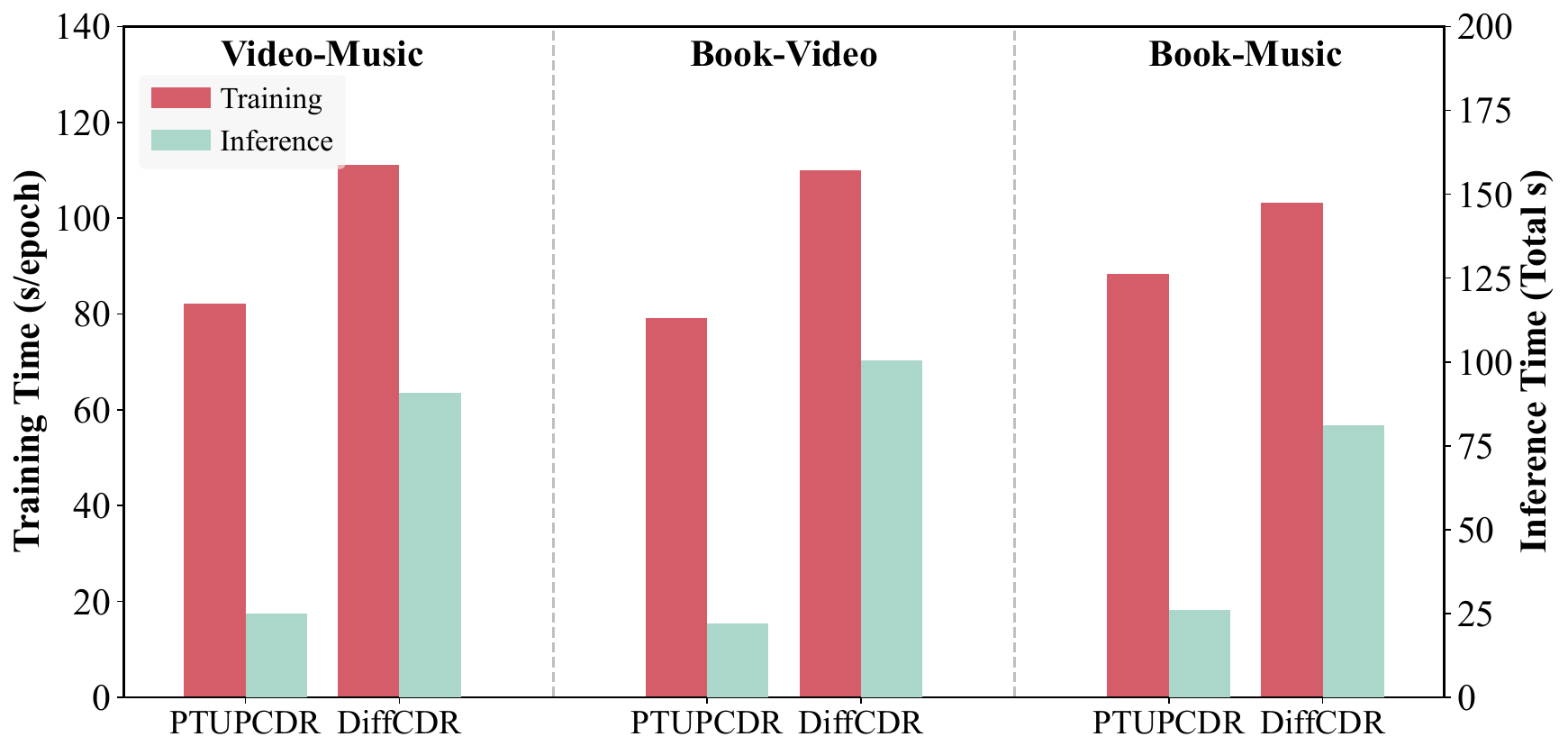}
        \captionsetup{labelfont=normalfont, textfont=normalfont}
        \caption{Cross-Domain}
        \label{fig:eff-cdr-merged}
    \end{subfigure}
    \caption{Efficiency Analysis. The left axis (Red bars) represents the Training Time per epoch, while the right axis (Blue hashed bars) denotes the Total Inference Time. Note that absolute inference time is dataset-dependent (e.g., number of users/items), thus we focus on within-dataset comparisons.}
    \label{fig:efficiency_all}
\end{figure*}

In the POI recommendation setting, Diff-POI consistently outperforms the non-diffusion baseline across all three datasets. A likely reason is that POI data couples spatial proximity with temporal dependency, and the progressive denoising sampling process can better integrate such spatiotemporal signals to produce more faithful recommendation scores than one-shot prediction.

In the cross-domain recommendation setting, DiffCDR achieves better results than the non-diffusion baseline on all three transfer tasks. This trend may be attributed to formulating target-domain representation recovery as conditional sampling guided by source-domain information, which supports representation alignment and mitigates cold-start effects; meanwhile, the conditioning design and sampling configuration can materially influence the final performance.

Across collaborative filtering, sequential, multimodal, POI, and cross-domain recommendation, the unified protocol yields directly comparable results within each scenario. 
Diffusion-based methods are often competitive with or superior to the selected non-diffusion baselines, although the gains vary across scenarios, datasets, and model designs.

\subsubsection{Computational Cost Analysis}

To provide a holistic view of the deployability of diffusion-based recommenders, we visualize the efficiency comparison in Figure~\ref{fig:efficiency_all}, reporting both the training time per epoch and the total inference time. Overall, diffusion-based recommenders tend to be more time-consuming than non-diffusion baselines, since they introduce additional computation from the multi-step denoising component and diffusion-related operations during training. Meanwhile, the cost varies noticeably across models and scenarios: when the diffusion process is coupled with a backbone with higher computational complexity, or when the scenario itself requires more complex encoders and larger candidate spaces, the training overhead typically becomes more pronounced. In contrast, comparatively lightweight diffusion designs can be more efficient, but they may still incur extra cost relative to conventional single-pass scoring models.
These results indicate that the empirical gains of diffusion-based methods often come with a non-negligible efficiency overhead, which is particularly important for large-scale or real-time recommendation settings. Therefore, beyond reporting ranking accuracy, a benchmark should also surface efficiency-related measurements to enable a more complete comparison and practical trade-off analysis. This observation further motivates future research on accelerating the sampling process, reducing the number of denoising steps, and developing more efficient parameterizations or distillation strategies to improve the deployability of diffusion-based recommenders in practical systems.

\subsection{Diffusion Configuration Analysis (RQ2)}

To systematically investigate the key factors influencing the performance of diffusion-based RSs and to provide meaningful guidance for future research, we conduct targeted analyses from the following three perspectives. 
In this section, we report experimental results in the collaborative filtering scenario for diffusion-based methods.

This analysis also validates the diffusion-aware diagnostic capability of Eval4DiRec. Unlike conventional recommendation frameworks that mainly compare final ranking metrics, Eval4DiRec explicitly exposes and records diffusion-specific configurations, including noise schedules, diffusion steps, noise distributions, and reverse-sampling settings. By varying these configurations under the same data processing, training, and evaluation protocol, we can examine how diffusion design choices affect recommendation quality and computational cost in a controlled manner.

\begin{figure}[ht]
    \centering
    \begin{subfigure}[b]{0.32\textwidth}
        \includegraphics[width=\linewidth]{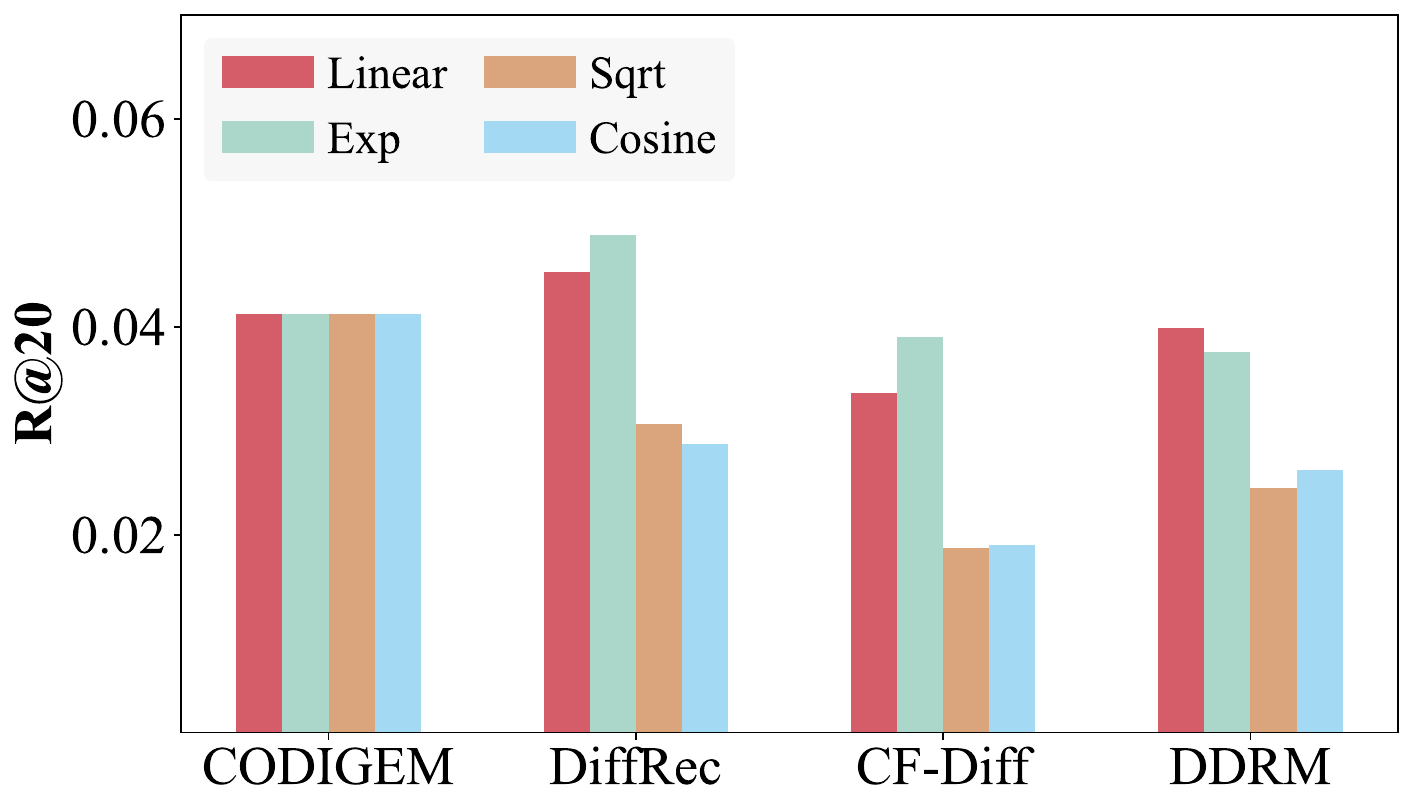}
        \captionsetup{labelfont=normalfont, textfont=normalfont}
        \caption{Yelp}
        \label{fig:schedule-yelp}
    \end{subfigure}
    \begin{subfigure}[b]{0.32\textwidth}
        \includegraphics[width=\linewidth]{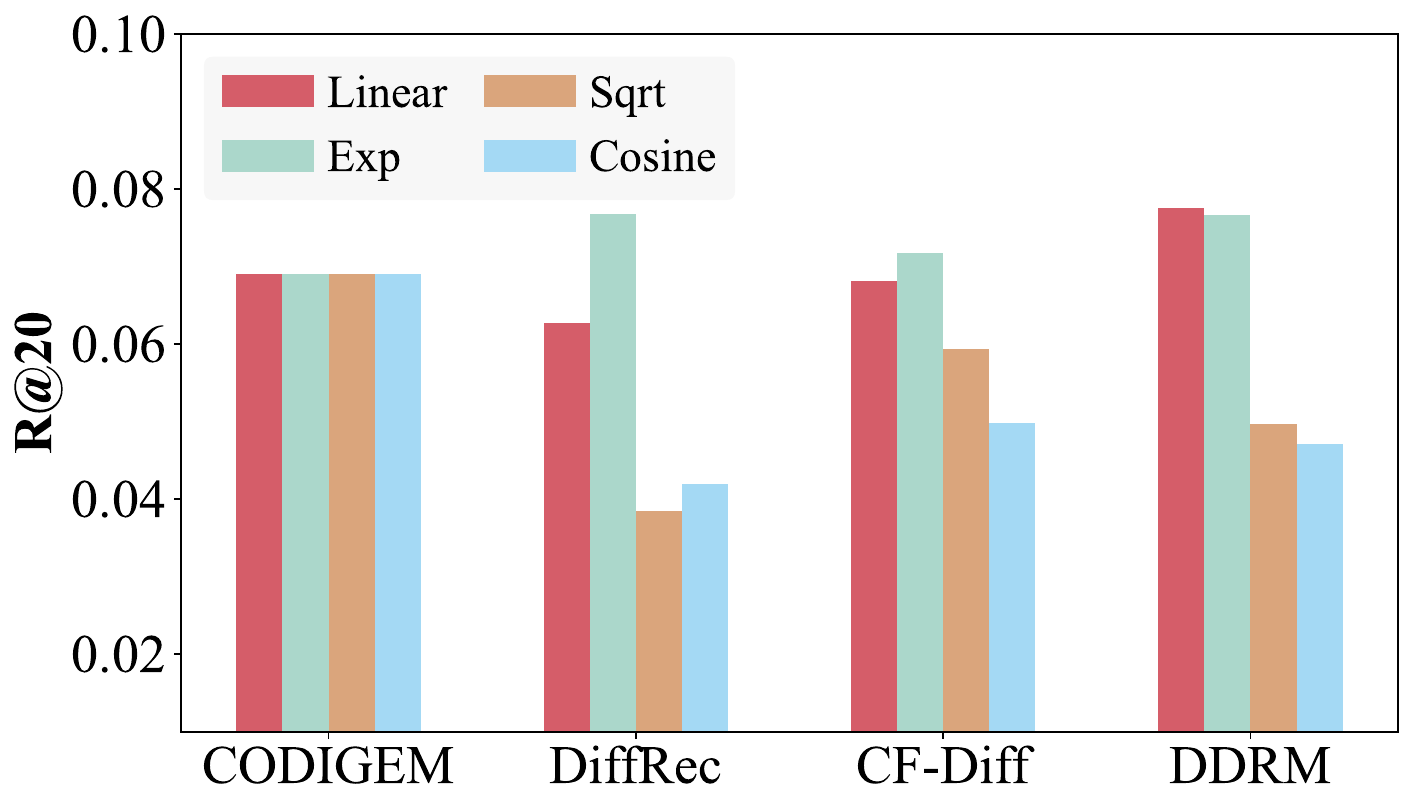}
        \captionsetup{labelfont=normalfont, textfont=normalfont}
        \caption{Beauty}
        \label{fig:schedule-beauty}
    \end{subfigure}
    \begin{subfigure}[b]{0.32\textwidth}
        \includegraphics[width=\linewidth]{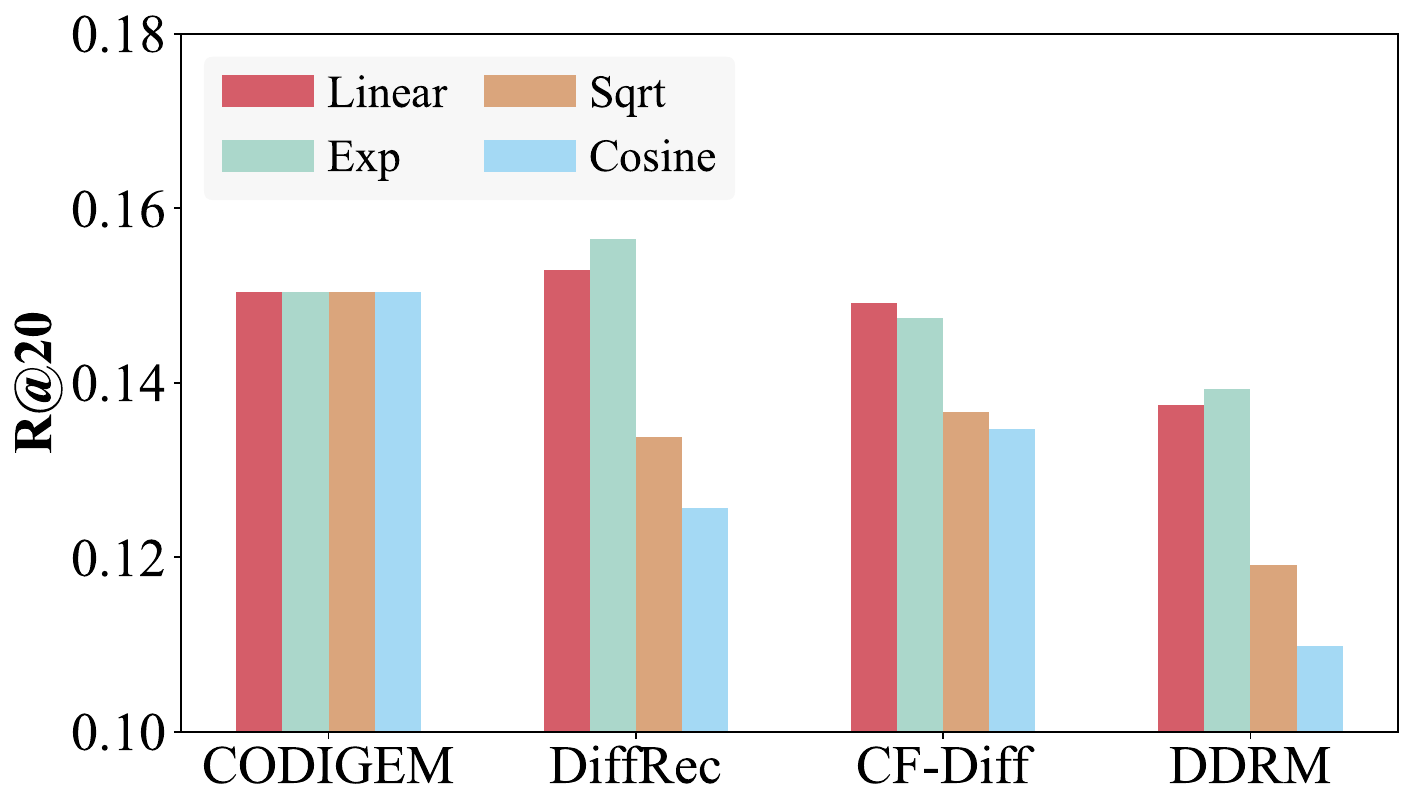}
        \captionsetup{labelfont=normalfont, textfont=normalfont}
        \caption{ML-1M}
        \label{fig:schedule-ml1m}
    \end{subfigure}
    \caption{Impact of noise schedule on diffusion-based recommendation performance on three datasets (Recall@20).}
    \label{fig:RQ2_schedule}
\end{figure}

\subsubsection{Noise Schedules and Recommendation Quality}

The noise schedule controls the noise magnitude injected at each diffusion timestep and therefore directly shapes the difficulty and behavior of the reverse denoising process~\cite{hang2024improved}. 
To examine its impact on recommendation performance, we compare four commonly used noise schedules under the same training and evaluation settings. 
As shown in Figure~\ref{fig:RQ2_schedule}, the linear and exponential schedules achieve the most stable and strongest performance overall, and the exponential schedule further surpasses the linear one on certain datasets. 
A plausible explanation is that these two schedules allocate noise across timesteps in a smoother and more structured manner, allowing the reverse process to progressively recover preference signals without either overwhelming personalization early on with overly aggressive perturbations or providing an insufficient learning signal at later stages due to too little corruption. 
By contrast, an ill-suited noise allocation may blur key personalized interaction patterns too quickly in the forward process, or make the denoising task overly easy, both of which can weaken the model’s learning and generation capability and ultimately lead to poorer ranking metrics. 
Overall, these results indicate that the noise schedule can substantially affect diffusion-based recommendation performance, suggesting that generic schedules may not always align well with the signal characteristics and learning objectives of recommendation tasks. 
This motivates future work on designing noise schedulers that are better tailored to recommendation scenarios.

\begin{figure}[ht]
    \centering
    \begin{subfigure}[b]{0.32\textwidth}
        \includegraphics[width=\linewidth]{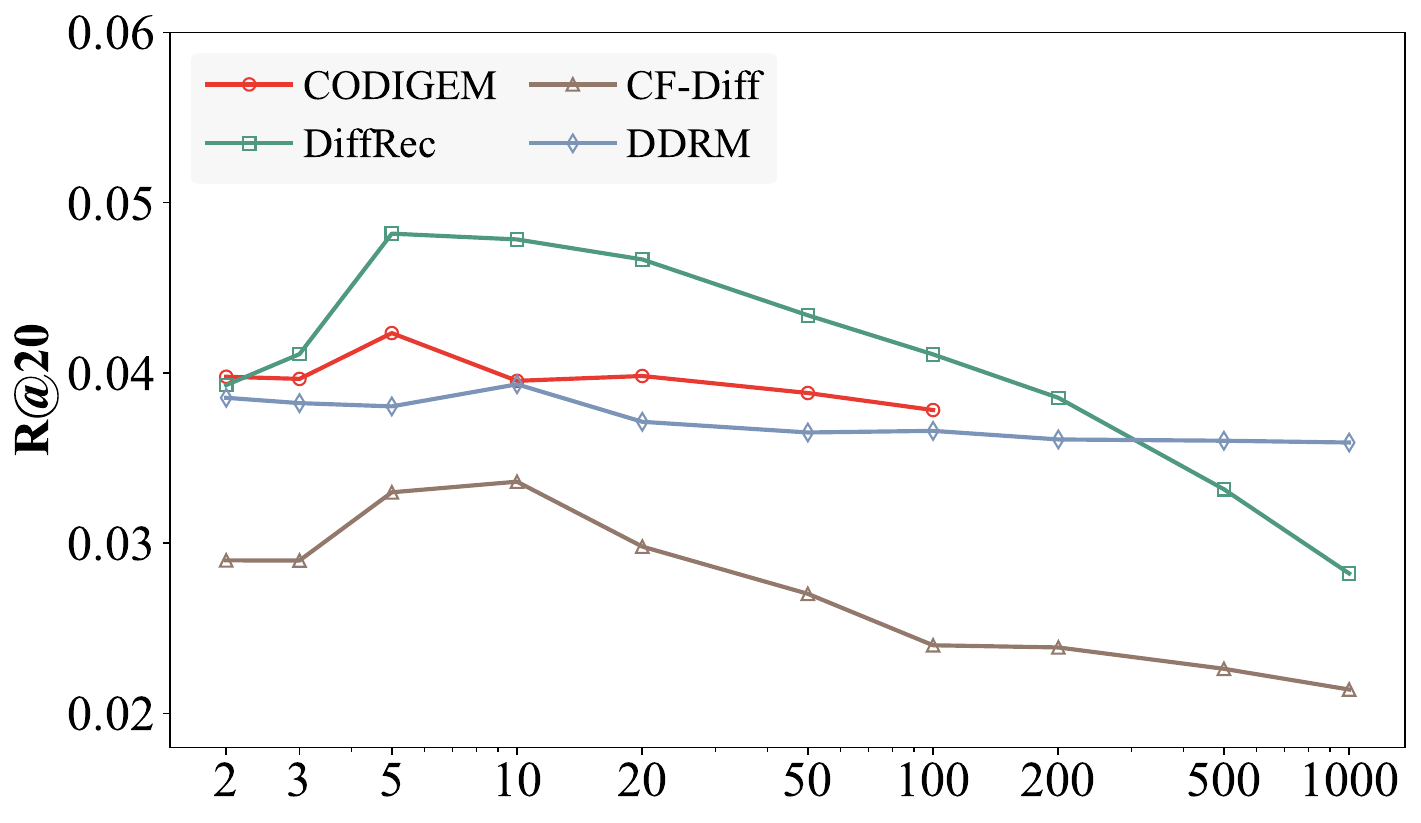}
        \captionsetup{labelfont=normalfont, textfont=normalfont}
        \caption{Yelp}
        \label{fig:timestep-yelp}
    \end{subfigure}
    \begin{subfigure}[b]{0.32\textwidth}
        \includegraphics[width=\linewidth]{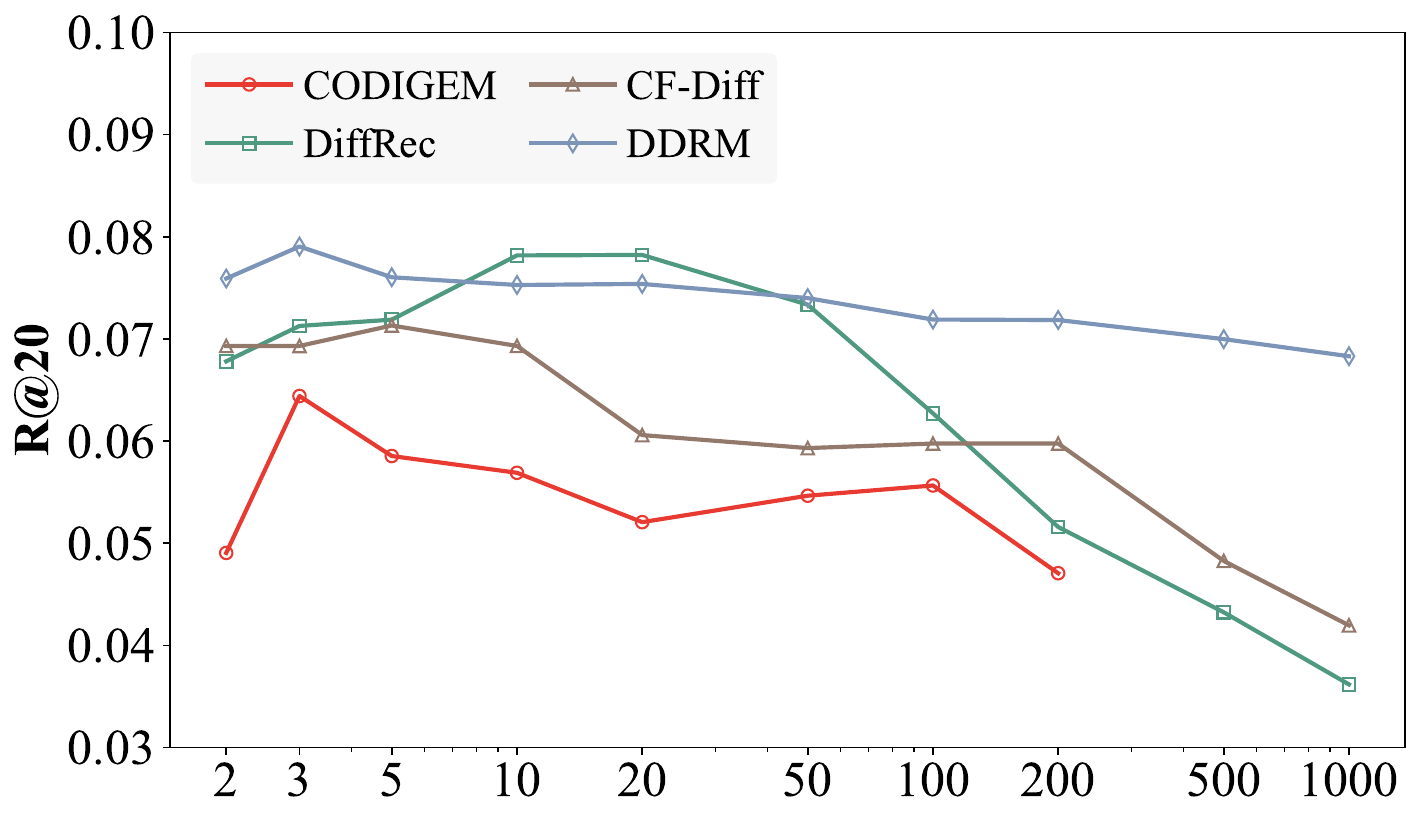}
        \captionsetup{labelfont=normalfont, textfont=normalfont}
        \caption{Beauty}
        \label{fig:timestep-beauty}
    \end{subfigure}
    \begin{subfigure}[b]{0.32\textwidth}
        \includegraphics[width=\linewidth]{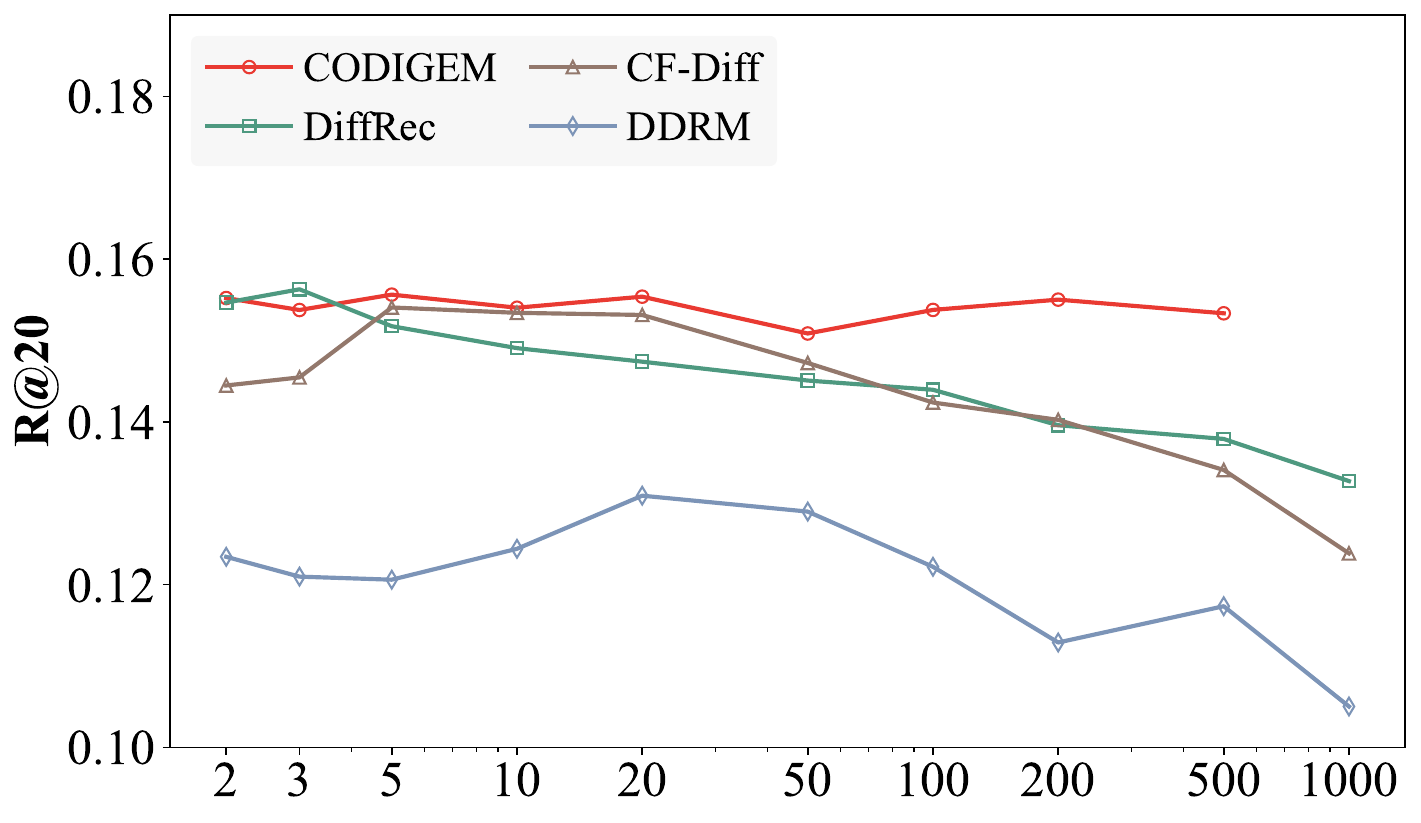}
        \captionsetup{labelfont=normalfont, textfont=normalfont}
        \caption{ML-1M}
        \label{fig:timestep-ml1m}
    \end{subfigure}
    \caption{Impact of diffusion steps on diffusion-based recommendation performance on three datasets (Recall@20).}
    \label{fig:RQ2_step}
\end{figure}

\subsubsection{Diffusion Steps: Quality--Cost Trade-off}

As part of our factor analysis, we examine how the number of diffusion steps affects the recommendation performance in the collaborative filtering scenario. 
The number of training diffusion steps defines the diffusion horizon used for the forward noising process and denoising objective during training. During inference, the forward process is not executed; instead, the independently configured number of inference steps determines the number of reverse denoising steps used to generate recommendation scores. 
As shown in Figure~\ref{fig:RQ2_step}, the best results are typically achieved with a relatively small number of steps; increasing the step count further often leads to diminishing returns and, in some cases, a noticeable performance drop. 
This trend can be attributed to the interaction between the recommendation signals and the multi-step denoising mechanism. 
Increasing the number of inference steps makes the sampling procedure longer and the reverse denoising process more computationally demanding, but it does not necessarily improve ranking quality. 
More prolonged perturbations can weaken discriminative preference cues, making it harder for the reverse process to recover ranking-relevant structure under limited supervision, and additional reverse steps may also accumulate errors during sampling, causing the generated signal to drift away from what ranking metrics favor. 
Overall, these results indicate that the number of diffusion steps should be treated as a key inference configuration that requires systematic tuning, rather than being increased by default, and that strong ranking performance often comes from carefully designed inference with a small number of steps.

\begin{figure}[ht]
    \centering
    \begin{subfigure}[b]{0.32\textwidth}
        \includegraphics[width=\linewidth]{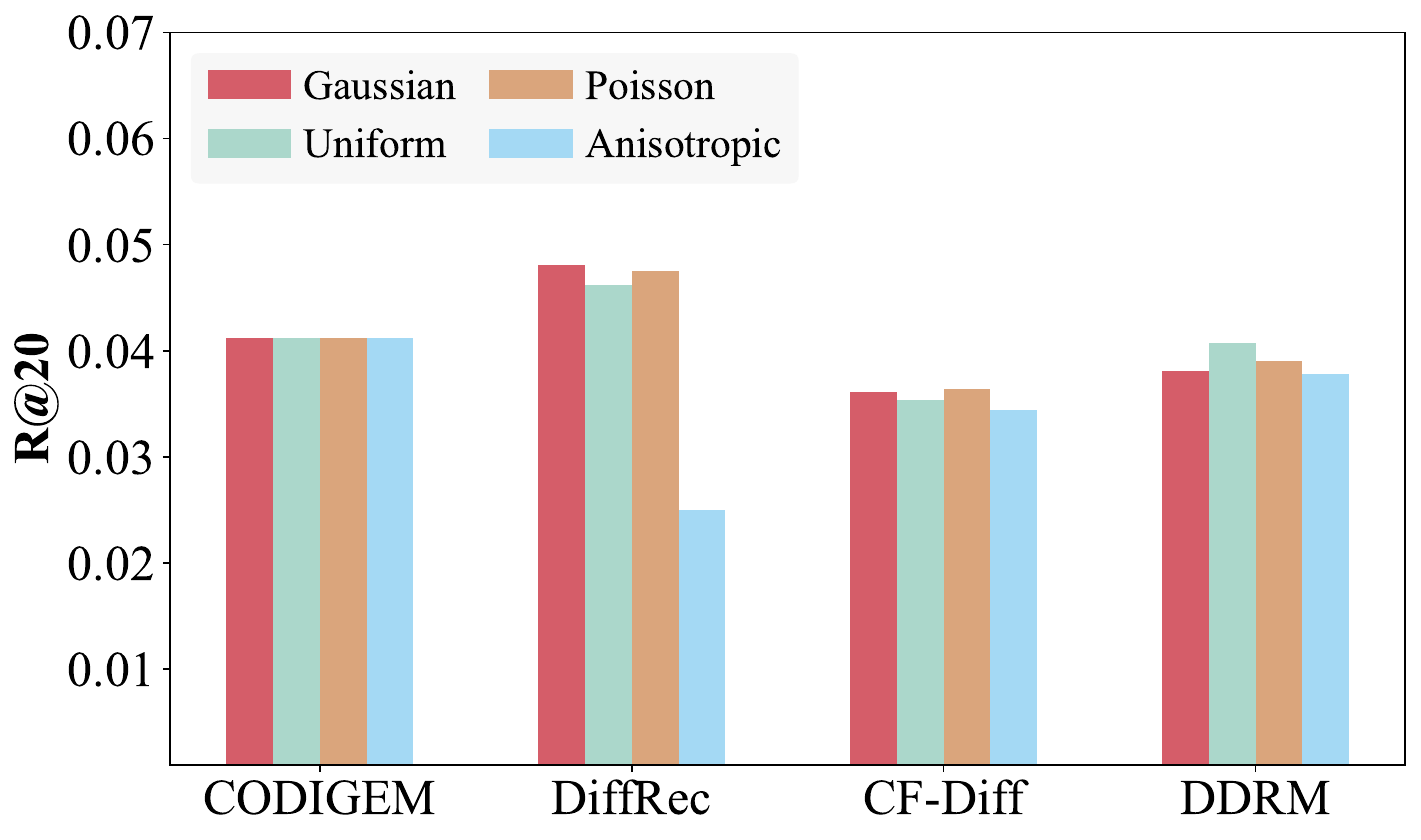}
        \captionsetup{labelfont=normalfont, textfont=normalfont}
        \caption{Yelp}
        \label{fig:distribution-yelp}
    \end{subfigure}
    \begin{subfigure}[b]{0.32\textwidth}
        \includegraphics[width=\linewidth]{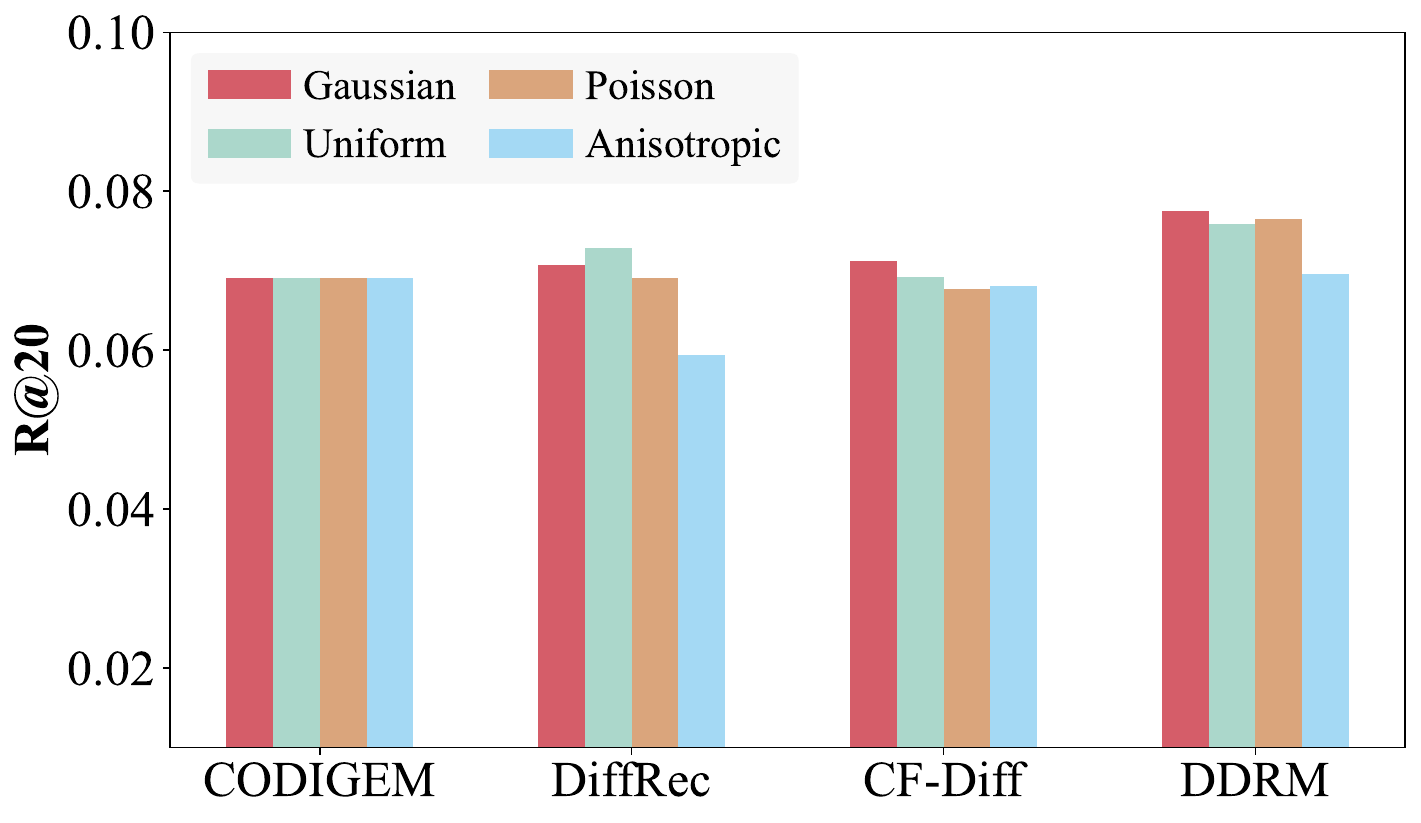}
        \captionsetup{labelfont=normalfont, textfont=normalfont}
        \caption{Beauty}
        \label{fig:distribution-beauty}
    \end{subfigure}
    \begin{subfigure}[b]{0.32\textwidth}
        \includegraphics[width=\linewidth]{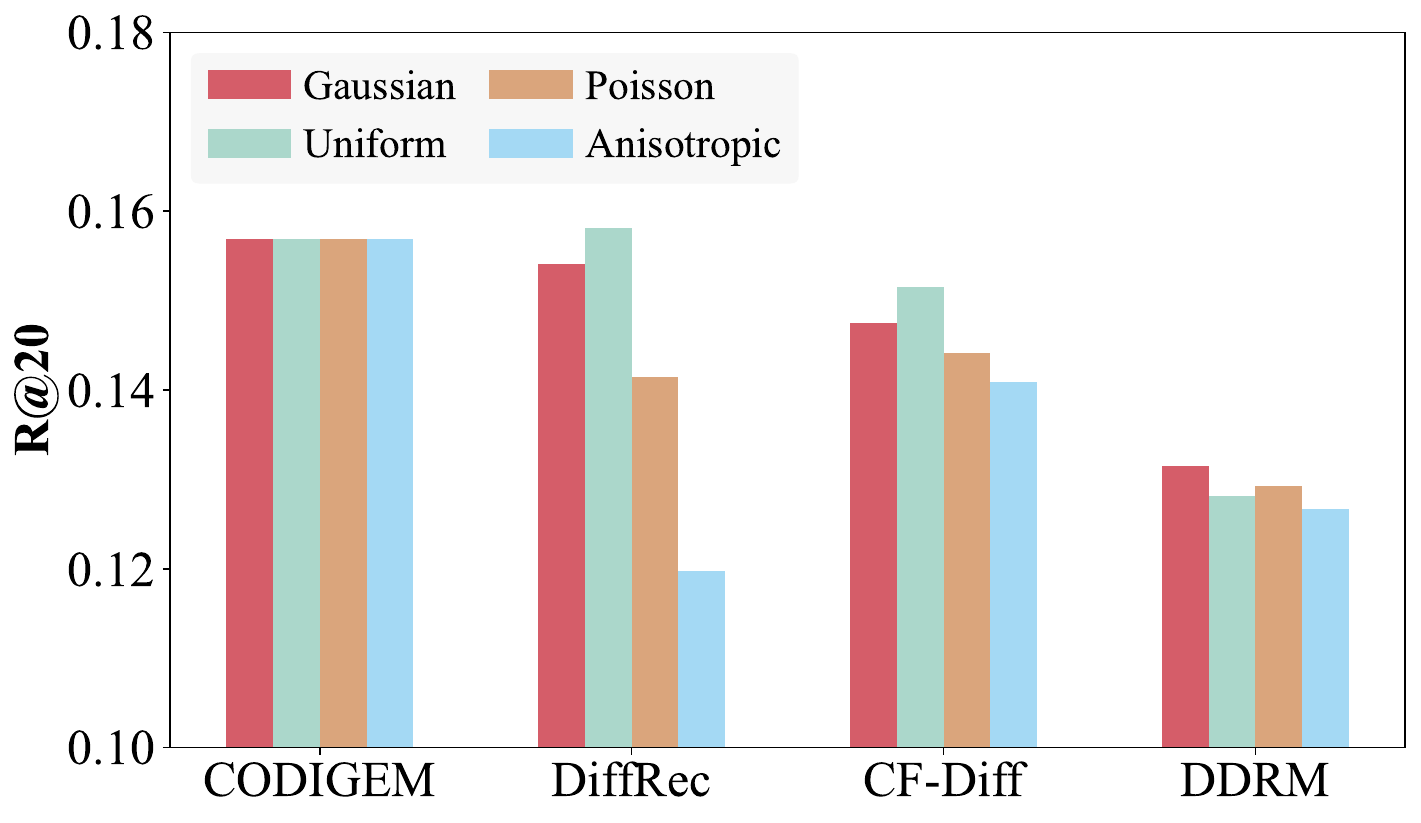}
        \captionsetup{labelfont=normalfont, textfont=normalfont}
        \caption{ML-1M}
        \label{fig:distribution-ml1m}
    \end{subfigure}
    \caption{Impact of noise distribution on diffusion-based recommendation performance on three datasets (Recall@20).}
    \label{fig:RQ2_distribution}
\end{figure}

\subsubsection{Noise Distributions and Recommendation Quality}

Noise distributions determine how the forward process perturbs recommendation signals, which in turn affects the difficulty and error characteristics of reverse denoising. Motivated by prior findings in the image domain, where alternatives from the exponential family have been explored~\cite{micheli2025diffusion}, we compare four noise distributions under the same training and evaluation settings: Gaussian, uniform, Poisson, and anisotropic Gaussian. As shown in Figure~\ref{fig:RQ2_distribution}, Gaussian noise provides the most consistent and stable performance across configurations, while alternative distributions do not dominate universally but can offer noticeable gains in particular datasets or inference settings. A plausible explanation is that recommendation data and representations can exhibit discrete, long-tailed, or anisotropic patterns that vary across scenarios; Gaussian perturbations tend to align with standard denoising assumptions and yield stable optimization, whereas uniform, Poisson, or anisotropic Gaussian noise may better match certain data regimes or feature geometries, leading to scenario-dependent benefits. Overall, the results indicate that noise distribution is a non-trivial design choice for diffusion-based recommendation and merits more systematic, task-aware investigation.

Taken together, the above results reveal that diffusion configurations jointly shape recommendation quality and computational cost by determining how recommendation signals are perturbed and recovered. Specifically, the noise schedule and noise distribution control the rate and form of information corruption, while the number of diffusion and reverse-sampling steps affects both the granularity of the recovery process and the required computational overhead. For sparse implicit-feedback data, a finer or longer diffusion process does not necessarily yield better recommendation performance. Excessive perturbation may weaken limited personalized signals, while additional reverse steps may accumulate recovery errors. This helps explain why recommendation quality changes non-monotonically with the number of diffusion steps, whereas inference cost generally increases with the length of the reverse process.

\subsection{Robustness Analysis of Diffusion-Based RSs (RQ3)}

Diffusion models excel at capturing complex data distributions, which may contribute to enhanced robustness in recommendation tasks. 
To systematically evaluate this property, we design two experiments targeting robustness under missing and noisy data conditions.

\begin{figure}[ht]
    \centering
    
    \begin{subfigure}[b]{0.32\textwidth}
        \includegraphics[width=\linewidth]{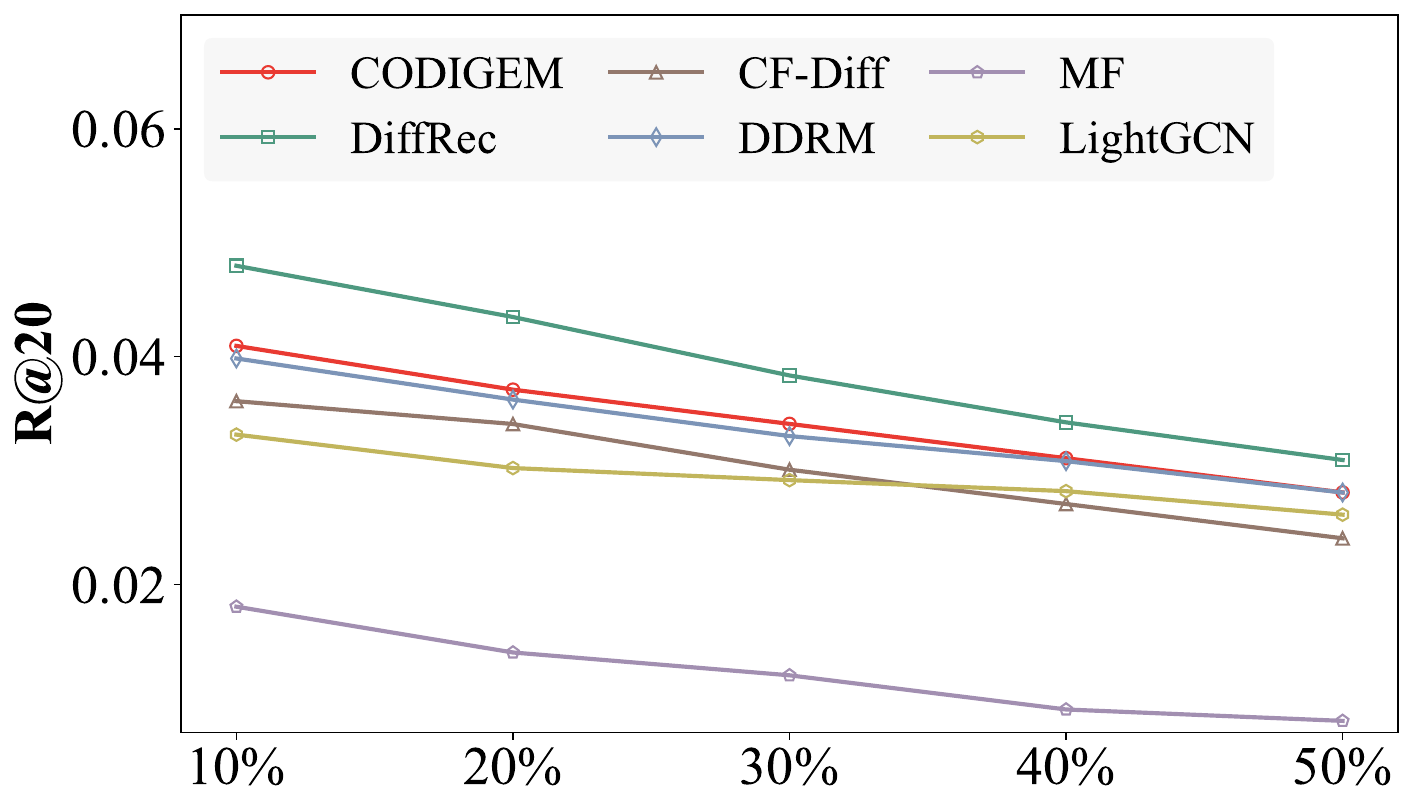}
        \captionsetup{labelfont=normalfont, textfont=normalfont}
        \caption{Yelp}
        \label{fig:sparsity-yelp}
    \end{subfigure}
    \begin{subfigure}[b]{0.32\textwidth}
        \includegraphics[width=\linewidth]{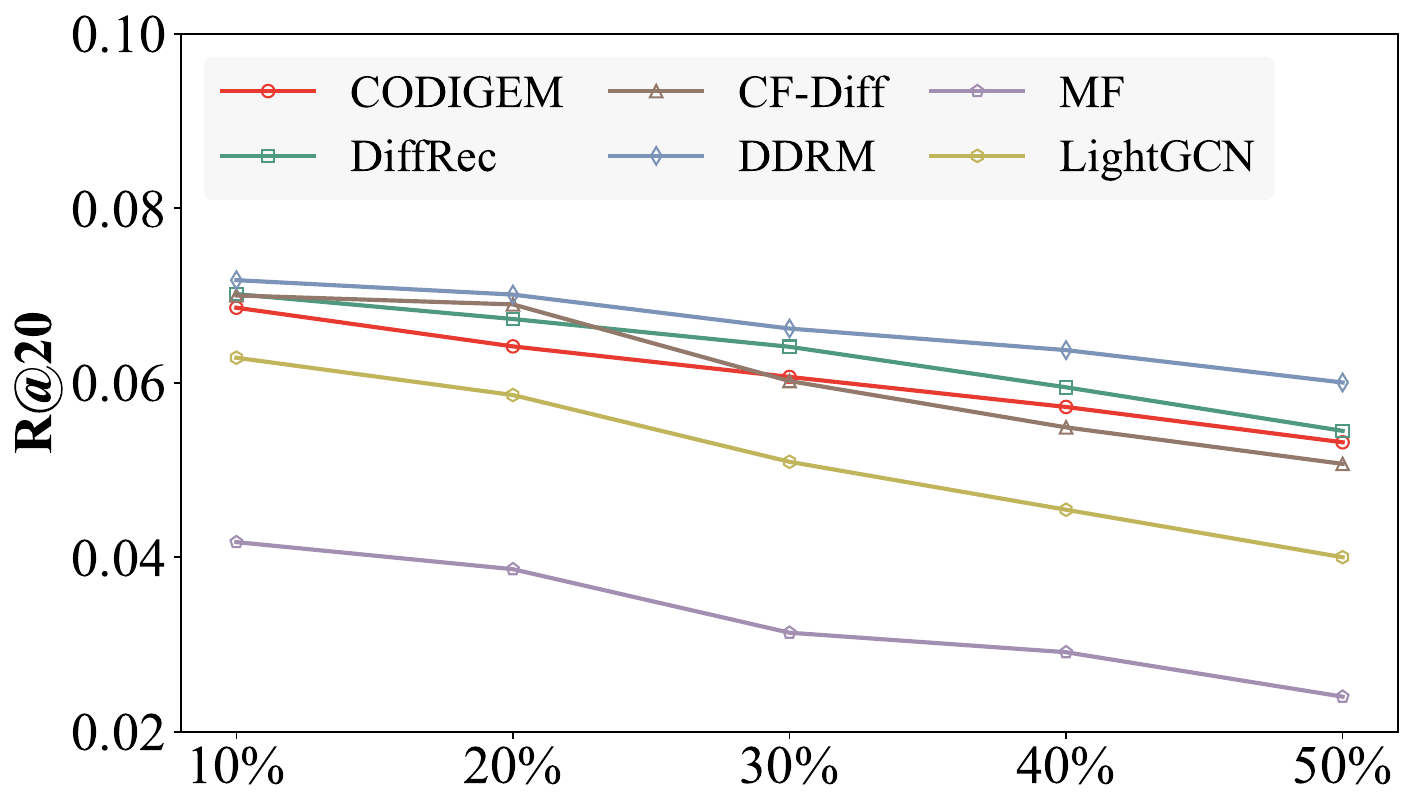}
        \captionsetup{labelfont=normalfont, textfont=normalfont}
        \caption{Beauty}
        \label{fig:sparsity-beauty}
    \end{subfigure}
    \begin{subfigure}[b]{0.32\textwidth}
        \includegraphics[width=\linewidth]{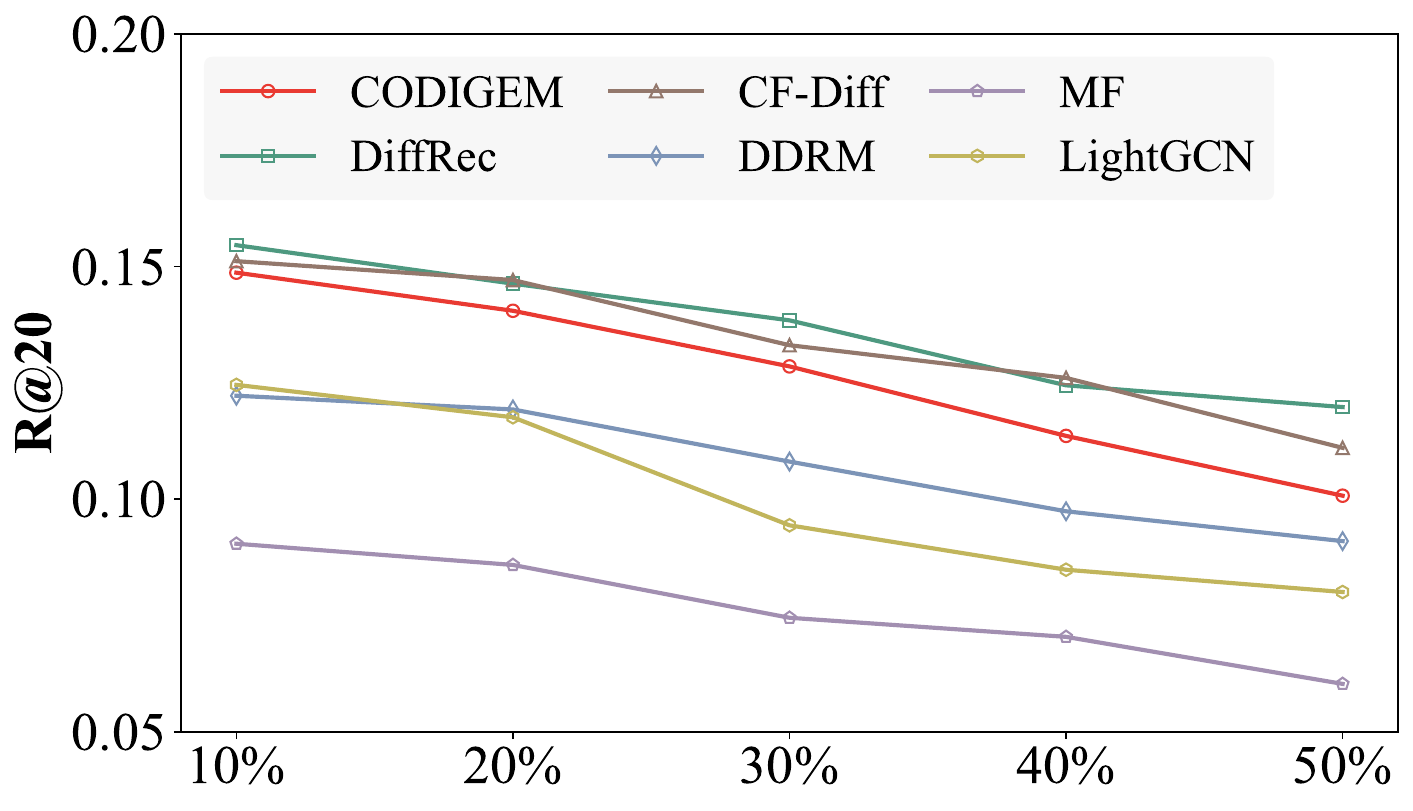}
        \captionsetup{labelfont=normalfont, textfont=normalfont}
        \caption{ML-1M}
        \label{fig:sparsity-ml1m}
    \end{subfigure}
    \caption{Robustness to missing data: performance under different missing ratios across three datasets (Recall@20).}
    \label{fig:RQ3_missing}
\end{figure}

\subsubsection{Robustness to Missing Data}

To evaluate robustness under missing training signals, we simulate increasing sparsity by randomly removing a portion of user--item interactions from the training set, with the removal ratio ranging from 10\% to 50\%. 
As shown in Figure~\ref{fig:RQ3_missing}, performance degrades for all methods as the available supervision decreases, which is expected given the reduced amount of preference evidence for learning. 
Notably, diffusion-based models exhibit a consistently milder drop than non-diffusion baselines, indicating stronger resilience to sparse supervision. 

A plausible explanation is that diffusion-based recommendation models are trained with an intrinsic perturbation and reconstruction objective. 
This learning paradigm encourages the model to recover preference-relevant latent signals from corrupted observations, which can partially alleviate representation degradation when interactions are missing. 
As a result, when supervision becomes limited, diffusion-based models tend to maintain more stable preference estimation, leading to better robustness in Top-$K$ ranking metrics.

\begin{figure}[ht]
    \centering
    
    \begin{subfigure}[b]{0.32\textwidth}
        \includegraphics[width=\linewidth]{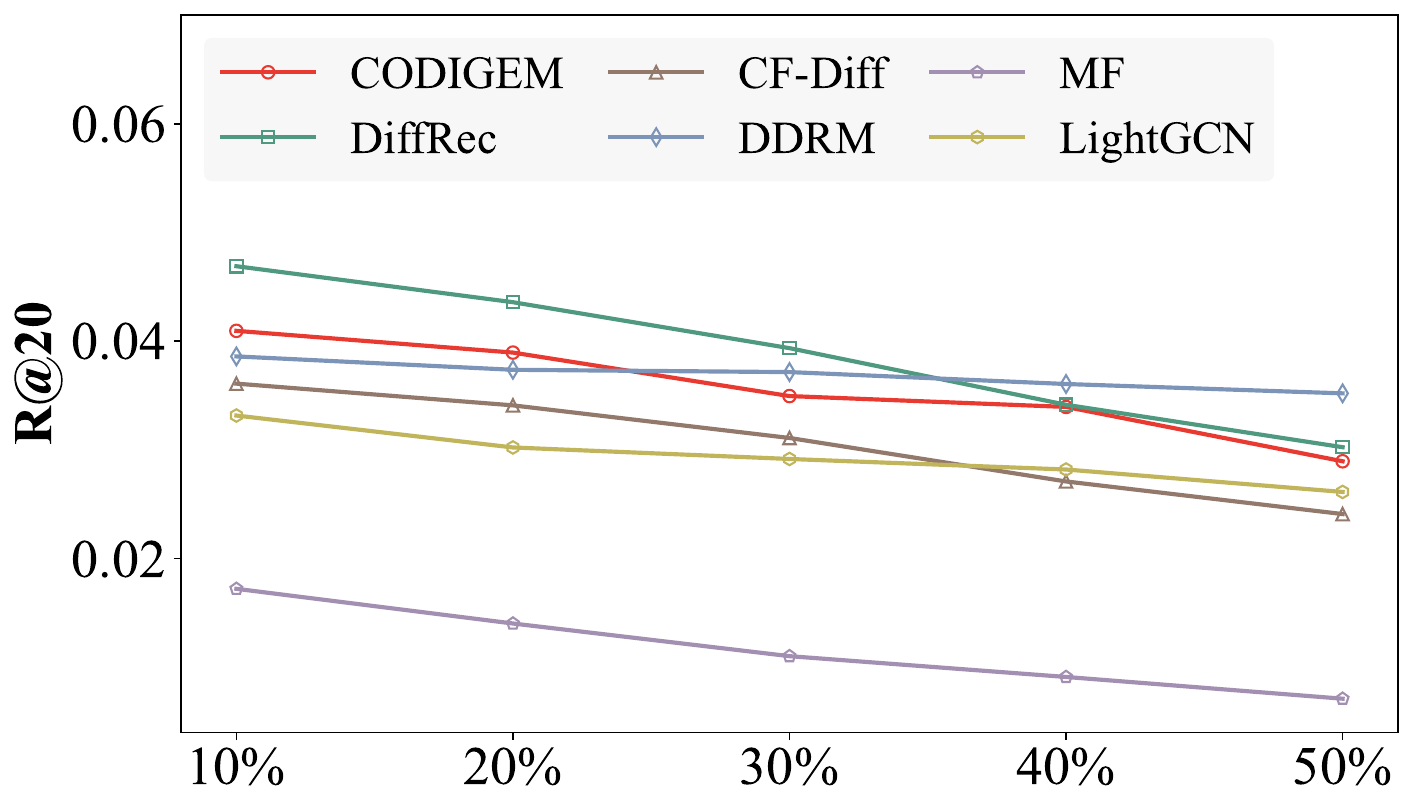}
        \captionsetup{labelfont=normalfont, textfont=normalfont}
        \caption{Yelp}
        \label{fig:noise-yelp}
    \end{subfigure}
    \begin{subfigure}[b]{0.32\textwidth}
        \includegraphics[width=\linewidth]{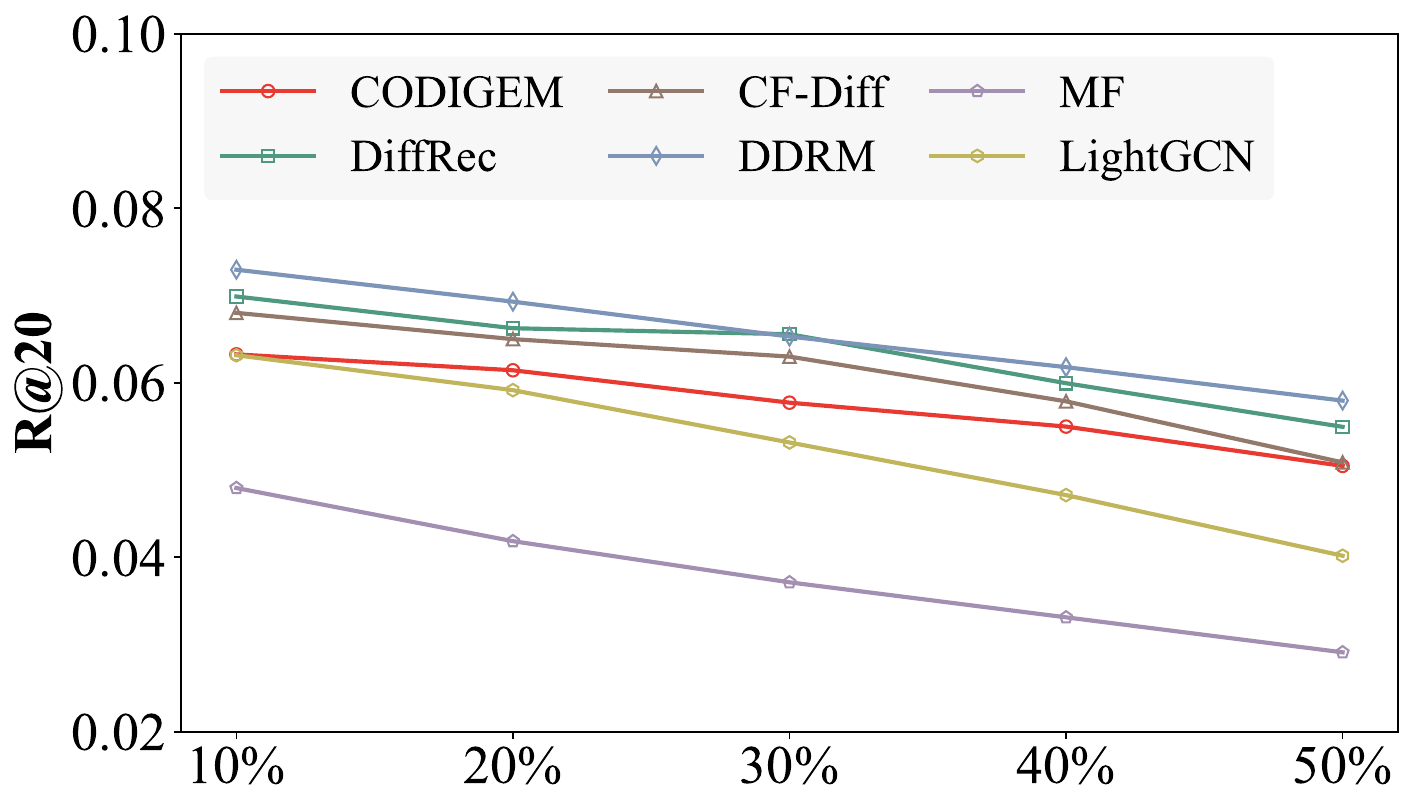}
        \captionsetup{labelfont=normalfont, textfont=normalfont}
        \caption{Beauty}
        \label{fig:noise-beauty}
    \end{subfigure}
    \begin{subfigure}[b]{0.32\textwidth}
        \includegraphics[width=\linewidth]{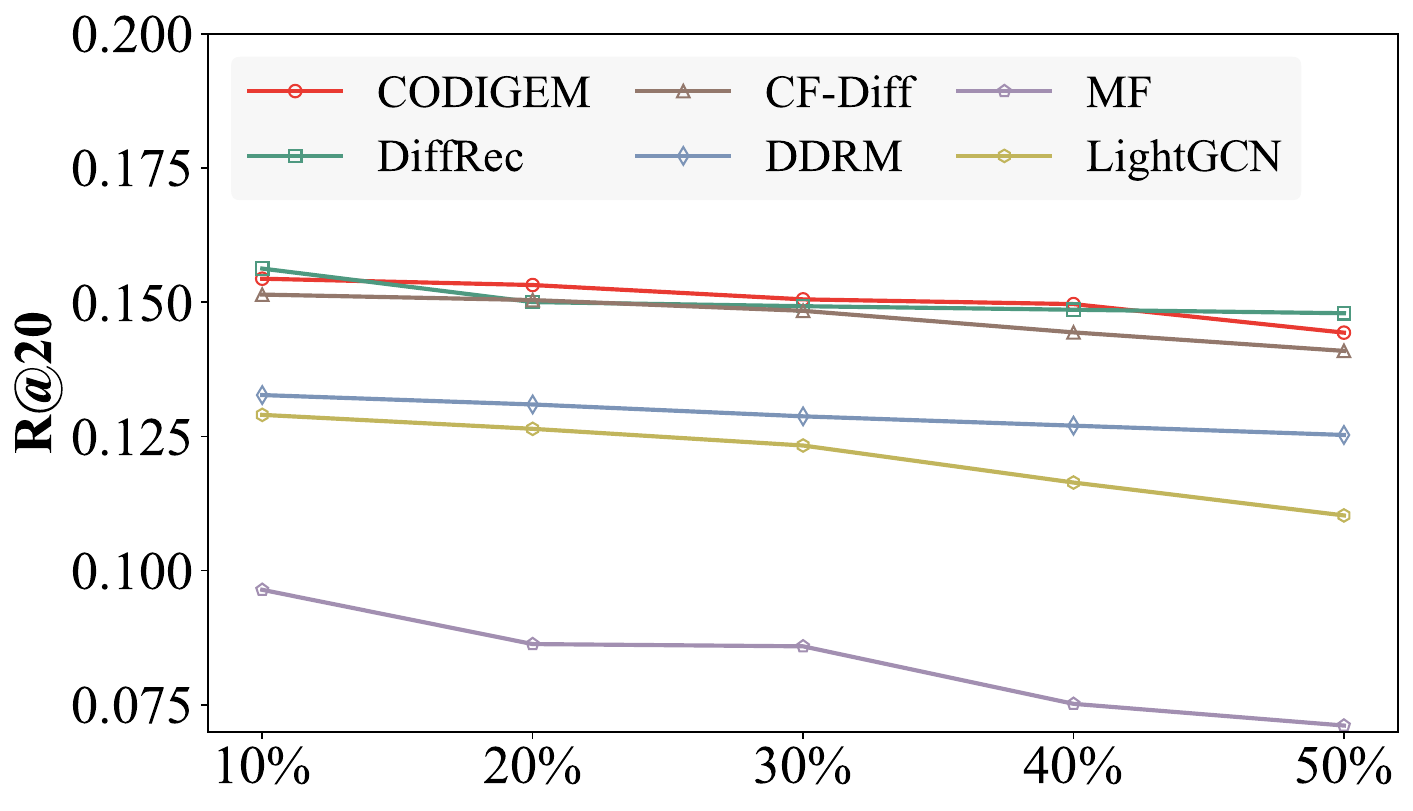}
        \captionsetup{labelfont=normalfont, textfont=normalfont}
        \caption{ML-1M}
        \label{fig:noise-ml1m}
    \end{subfigure}
    \caption{Robustness to noisy data: performance under different noise ratios across three datasets (Recall@20).}
    \label{fig:RQ3_noise}
\end{figure}

\subsubsection{Robustness to Noisy Data}

Beyond missing data, real-world implicit feedback is often contaminated by noise, such as accidental clicks, exposure bias, or logging errors. 
To assess robustness to noisy supervision, we inject randomly sampled negative items into the training set at the same rate, with the noise ratio also varying from 10\% to 50\%. 
Figure~\ref{fig:RQ3_noise} shows that increasing noise consistently harms ranking performance across all methods, while diffusion-based models again exhibit noticeably less severe degradation than non-diffusion baselines.

This behavior can be attributed to the denoising-oriented modeling principle of diffusion methods. 
By learning to reverse controlled perturbations, diffusion-based recommendation models are encouraged to capture stable and structured preference patterns rather than overfitting spurious correlations introduced by random noise. 
Consequently, under corrupted supervision, the denoising generation process can provide an implicit error-correction effect, mitigating the negative impact of noisy interactions on the final ranking quality.

The robustness results provide a complementary perspective. Diffusion-based recommenders often exhibit smaller performance degradation under missing or noisy interactions, which may be related to the regularizing effect of learning to recover stable preference structures from perturbed inputs. However, this observation should not be interpreted as evidence that diffusion models are inherently robust to all forms of real-world noise. The artificial perturbations used in diffusion training differ from accidental clicks, exposure bias, logging errors, and non-randomly missing interactions in practical recommendation systems. Therefore, the robustness benefit depends on how well the learned denoising process matches the actual corruption patterns in the data.

Taken together, the analyses show that recommendation quality, computational cost, and robustness should be considered jointly when evaluating diffusion-based recommenders. Strong performance on clean data does not necessarily imply low inference cost or limited degradation under data corruption. Accordingly, diffusion-based models and configurations should not be selected solely according to a single accuracy metric. Instead, they should be evaluated under a unified protocol that jointly considers recommendation quality, computational efficiency, and robustness, with model-specific diffusion and sampling settings fully reported.

\begin{figure}[ht]
    \centering
    \begin{subfigure}[b]{0.32\textwidth}
        \includegraphics[width=\linewidth]{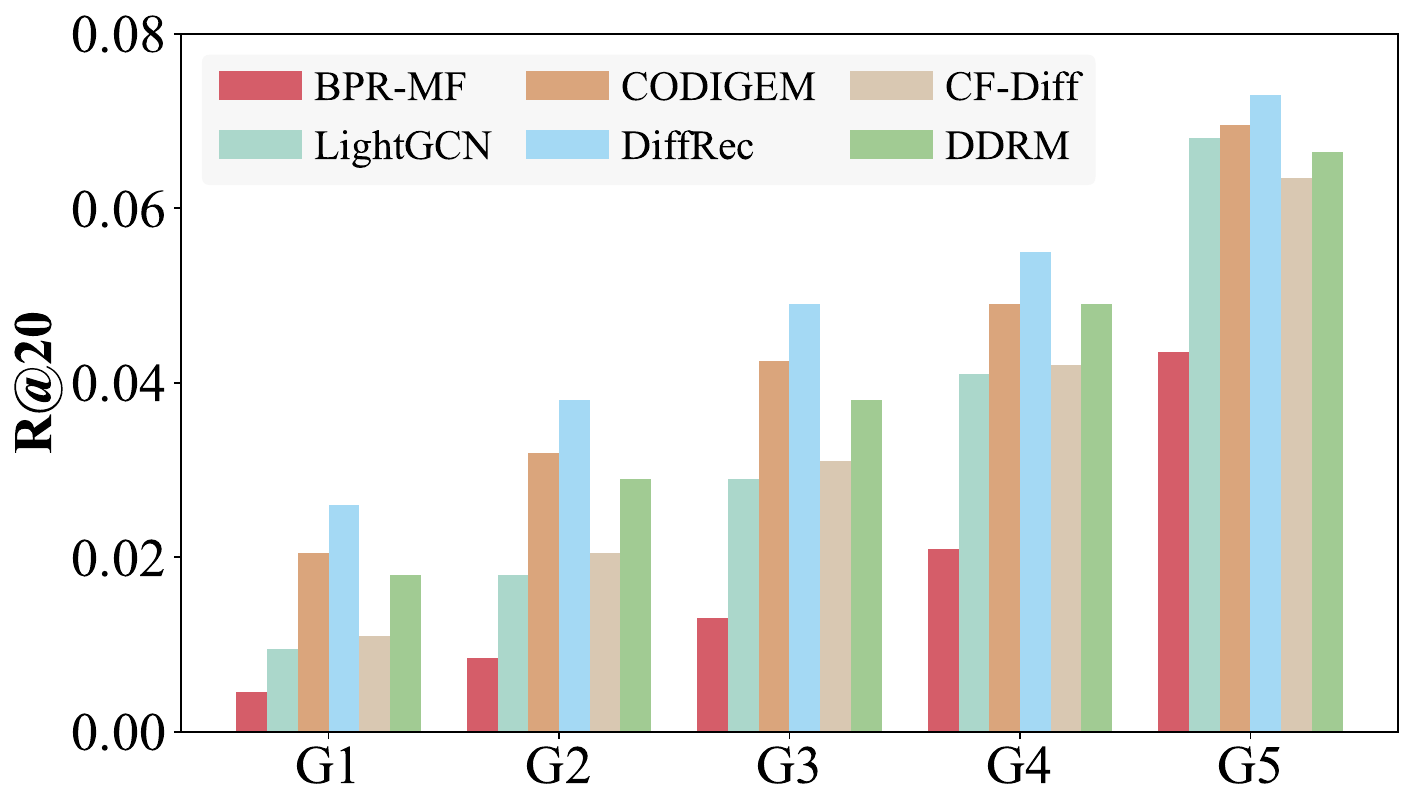}
        \captionsetup{labelfont=normalfont, textfont=normalfont}
        \caption{Yelp}
        \label{fig:longtail-yelp}
    \end{subfigure}
    \begin{subfigure}[b]{0.32\textwidth}
        \includegraphics[width=\linewidth]{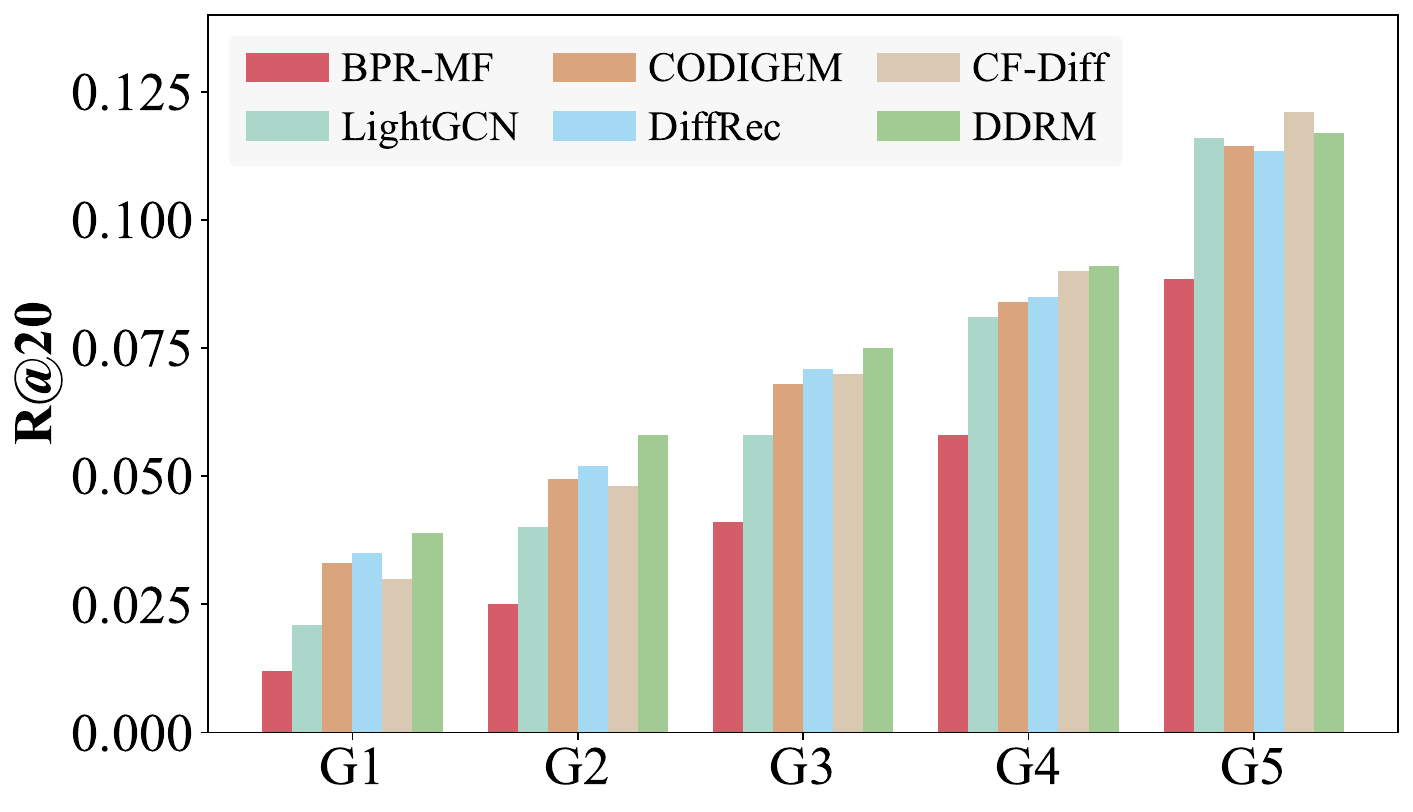}
        \captionsetup{labelfont=normalfont, textfont=normalfont}
        \caption{Beauty}
        \label{fig:longtail-beauty}
    \end{subfigure}
    \begin{subfigure}[b]{0.32\textwidth}
        \includegraphics[width=\linewidth]{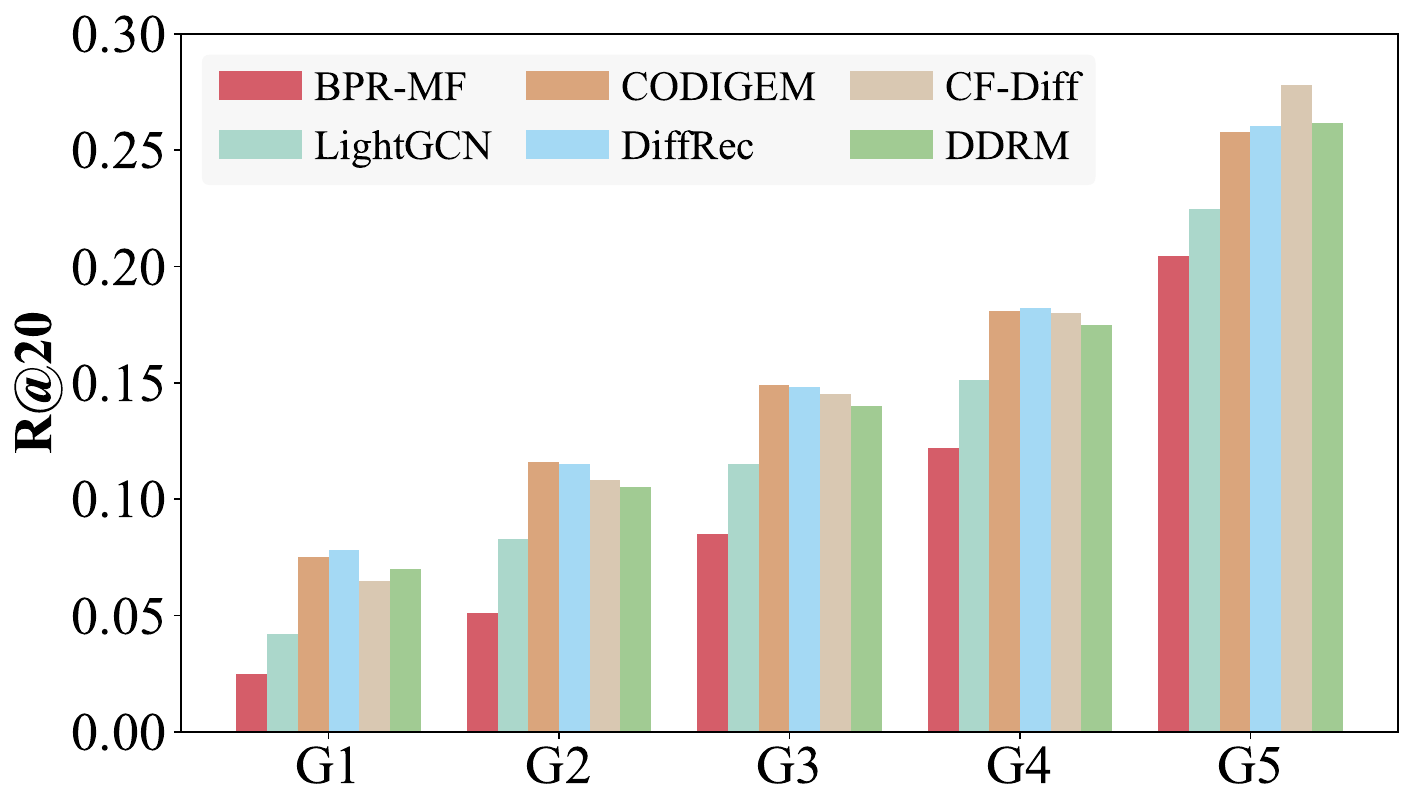}
        \captionsetup{labelfont=normalfont, textfont=normalfont}
        \caption{ML-1M}
        \label{fig:longtail-ml1m}
    \end{subfigure}
    \caption{Long-tail item analysis across different popularity groups on three datasets (Recall@20).}
    \label{fig:RQ4_longtail}
\end{figure}

\subsection{Analysis of Long-tail Items (RQ4)}
As illustrated in Figure~\ref{fig:RQ4_longtail}, we partition items into popularity groups according to their interaction frequency in the training data, ordered from head to long-tail. 
A clear disparity can be observed for non-diffusion baselines across these groups: they achieve relatively strong performance on head items with abundant interactions, but their accuracy drops substantially as item popularity decreases, and recall becomes particularly weak for long-tail groups. 
In contrast, diffusion-based methods exhibit a noticeably milder degradation on less popular items and produce smoother performance curves, indicating that they can maintain more usable ranking signals under weaker supervision and deliver more consistent gains in the long-tail region.

A plausible explanation is that the two families of methods exploit training signals in different ways. 
Non-diffusion models typically rely on directly observed interactions to learn item representations and scoring functions. 
When an item has few interactions, the effective supervision for parameter updates becomes limited, and the model can be dominated by high-frequency items, which often leads to inferior representation quality and reduced discriminability for long-tail items. 
Diffusion-based methods, in comparison, optimize a progressive perturbation and recovery objective. 
During training, the model repeatedly observes the same signal under different noise levels and learns to recover consistent structure from incomplete or corrupted inputs. 
This learning mechanism can implicitly strengthen sparse supervision, giving the model more opportunities to capture latent associations between users and long-tail items, and thereby alleviating performance collapse in tail groups. 
We also observe heterogeneous gains among diffusion-based methods. Some models yield larger improvements on long-tail groups, while others preserve stronger advantages on head items, suggesting that diffusion does not automatically bring uniform benefits across all popularity levels. 
The actual effect depends on diffusion configurations, the sampling procedure, and how the diffusion process is coupled with the recommendation backbone. 
Overall, this long-tail analysis not only highlights the potential of diffusion-based recommendation under low-frequency supervision, but also motivates future work to better balance tail coverage and head accuracy, and to develop diffusion modeling and sampling strategies that are more aligned with the characteristics of recommendation data.

\section{Conclusion and Future Work}
Diffusion-based recommendation has advanced rapidly in recent years, yet empirical studies remain difficult to reproduce and compare fairly due to inconsistent practices in data processing, dataset splitting, metric implementation, and inference settings. 
To address these issues, we propose Eval4DiRec, a unified and reproducible evaluation framework for diffusion-based recommender systems. 
Eval4DiRec provides an end-to-end standardized pipeline with a modular implementation, enabling consistent benchmarking under unified protocols. 
Based on this framework, we integrate 14 representative diffusion-based model implementations across five recommendation scenarios and conduct systematic experiments on ranking performance, efficiency, diffusion configurations, robustness under missing and noisy interactions, and long-tail behavior.
The results demonstrate that diffusion-based recommenders can be competitive, but their gains are scenario- and configuration-dependent and often come with additional computational cost. This variability reinforces the need for standardized and transparent evaluation.

Although Eval4DiRec provides standardized benchmarking for diffusion-based recommender systems, its current coverage remains representative rather than exhaustive. Emerging paradigms, such as bridge-based generation and spectral-space diffusion, may involve substantially different state spaces, training objectives, and conditioning mechanisms that have not yet been incorporated into the current framework. In addition, the existing evaluation suite mainly focuses on ranking quality, computational time, robustness, and long-tail performance, while reliability-oriented dimensions, such as diversity, calibration, and fairness, have not yet been systematically evaluated.

Future work will address these limitations from both architectural and evaluation perspectives. At the architectural, we plan to further decouple the diffusion process into reusable components, including state representations, perturbation kernels, noise schedulers, conditioning mechanisms, training objectives, and reverse samplers. This component-level design will facilitate the integration of emerging diffusion architectures without requiring substantial redesign of the existing training and evaluation pipeline. At the evaluation, we will extend the benchmark beyond accuracy-centered comparisons toward a more comprehensive analysis of the trade-offs among recommendation quality, efficiency, and robustness across different data scales. We will also investigate reduced-step inference, lightweight samplers, and knowledge distillation to lower deployment costs while preserving recommendation quality. Finally, we will strengthen reproducibility and long-term maintainability through more complete configuration and environment records and community-driven model integration.

\begin{acks}
This work was supported by National Natural Science Foundation of China (No. 62376130), Program of New Twenty Policies for Universities of Jinan (Grant No.202333008), and the Pilot Project for Integrated Innovation of Science, Education, and Industry of Qilu University of Technology (Shandong Academy of Sciences) (No.2025ZDZX01).
\end{acks}

\bibliographystyle{ACM-Reference-Format}
\bibliography{main}

\end{document}